\documentclass[12pt]{article}

\usepackage{bm}
\usepackage{amsmath,amssymb}
\allowdisplaybreaks

\usepackage{wick}

\begin{document}
\baselineskip=16.0pt plus 0.2pt minus 0.1pt

%\baselineskip=16.8pt plus 0.2pt minus 0.1pt

%%%%%%%%%%%%%%%%%%%%%%%%%%%%%%%%%%%%%%%%%%%%%%%%%%%%%%%%%%%%%%%%
\makeatletter
\@addtoreset{equation}{section}
\renewcommand{\theequation}{\thesection.\arabic{equation}}
\renewcommand{\thefootnote}{\fnsymbol{footnote}}

\newcommand{\Tr}{\mathop{\textrm{Tr}}}
\newcommand{\Po}{\mathop{\textrm{P}}}
\newcommand{\Xw}{\mathop{{\cal X}}}
\newcommand{\N}{{\cal N}}
\newcommand{\W}{{\cal W}}
\newcommand{\D}{{\cal D}}

\newcommand{\F}{{\cal F}}
\newcommand{\A}{{\cal A}}
\newcommand{\E}{{\cal E}}
\newcommand{\Op}{{\cal O}}
\newcommand{\ds}{\displaystyle}
\newcommand{\s}{\sigma}
\newcommand{\VEV}[1]{\left\langle #1\right\rangle}
\newcommand{\bra}[1]{\left\langle #1\right|}
\newcommand{\ket}[1]{\left| #1\right\rangle}
\newcommand{\braket}[2]{\langle #1\vert #2\rangle}
\newcommand{\fDrv}[2]{\frac{\delta #1}{\delta #2}}
\newcommand{\p}{\partial}
\newcommand{\ol}[1]{\overline{#1}}

% \slashchar puts a slash through a character to represent contraction
% with Dirac matrices. Use \not instead for negation of relations, and
% use \hbar for hbar.
\def\slashchar#1{\setbox0=\hbox{$#1$} % set a box for #1
  \dimen0=\wd0 % and get its size
  \setbox1=\hbox{/} \dimen1=\wd1 % get size of /
  \ifdim\dimen0>\dimen1 % #1 is bigger
  \rlap{\hbox to \dimen0{\hfil/\hfil}} % so center / in box
  #1 % and print #1
  \else % / is bigger
  \rlap{\hbox to \dimen1{\hfil$#1$\hfil}} % so center #1
  / % and print /
  \fi}
\newcommand{\nd}{$\slashchar{\delta}$}
\newcommand{\DS}{\slashchar{D}}

\newcommand{\refadd}{{\bf [Refs]}\ }

\makeatother
%%%%%%%%%%%%%%%%%%%%%%%%%%%%%%%%%%%%%%%%%%%%%%%%%%%%%%%%%%%%%%%%%

\begin{titlepage}
\title{
\hfill\parbox{4cm}
{\normalsize KUNS-1990\\{\tt hep-th/0510150}}\\
\vspace{1cm}
{\bf Loop Equation in $\bm{D=4}$, ${\bm{\N = 4}}$ SYM
 and \\String Field Equation on
${\rm \bf AdS}_{\bm 5} \times {\rm \bf S}^{\bm 5}$
}
}
\author{
Hiroyuki {\sc Hata}\thanks{
{\tt hata@gauge.scphys.kyoto-u.ac.jp}}
\
and
\
Akitsugu {\sc Miwa}\thanks{
{\tt akitsugu@gauge.scphys.kyoto-u.ac.jp}}
\\[15pt]
{\it Department of Physics, Kyoto University, Kyoto 606-8502, Japan}
}
\date{\normalsize October, 2005}
\maketitle
\thispagestyle{empty}
\vspace{2cm}
\begin{abstract}
\normalsize
We consider the loop equation in four-dimensional $\N = 4$ SYM,
which is a functional differential equation for the Wilson loop $W(C)$
and expresses the propagation and the interaction of the
string $C$. Our $W(C)$ consists of the scalar and the gaugino fields
as well as the gauge field.
The loop $C$ is specified by six bosonic coordinates $y^i(s)$
and two fermionic coordinates $\zeta(s)$ and $\eta(s)$
besides the four-dimensional spacetime coordinates $x^\mu(s)$.
We have successfully determined, to quadratic order in $\zeta$ and
$\eta$, the parameters in $W(C)$ and the loop differential operator
so that the equation of motion of SYM can be correctly reproduced to
give the non-linear term of $W(C)$.
We extract the most singular and linear part of our loop equation and
compare it with the Hamiltonian constraint of the string propagating
on ${\rm AdS}_5 \times {\rm S}^5$ background.

\end{abstract}
\end{titlepage}

\vfill
\newpage

\section{Introduction}
\label{introduction}

The large $N$ dualities between string theories and gauge
theories are of great importance for the understanding of the both
theories. The ${\rm AdS}/{\rm CFT}$ correspondence
\cite{Maldacena:1997re,Gubser:1998bc,Witten:1998qj,Aharony:1999ti,
D'Hoker:2002aw}
is one of the most interesting examples of such dualities. On the
string theory side of this correspondence, we consider the type IIB
superstring on ${\rm AdS}_5 \times {\rm S}^5$ geometry.
On the other hand, the corresponding gauge theory is the
four-dimensional $\N = 4$ SU($N$) super Yang-Mills theory (SYM).
Since this correspondence was first conjectured, a great deal
of aspects of it have been studied.
Among them, the correspondences associated with the Wilson
loop operator of SYM
\cite{Rey:1998ik,Maldacena:1998im,Drukker:1999zq,Drukker:1999gy,
Polyakov:2000ti,Polyakov:2000jg,Semenoff:2002kk}
seem to be very important, because the proposed counterpart in the
string theory side is nothing but the fundamental string.
The standard argument of the correspondence begins with considering
the following Wilson loop operator $W(C)$ defined on the loop
$C$:
\begin{equation}
W(C)=\Tr\Po\exp\left(
i\int_0^l ds\Bigl(A_\mu(x(s))\dot x^\mu(s)+A_{3+i}(x(s))\dot y^i(s)
\Bigr)\right),
\label{intsuperW}
\end{equation}
where $A_\mu$ ($\mu = 0 , \ldots ,3$)
and $A_{3+i}$ ($i = 1, \ldots , 6$)
are the gauge field and the six scalar fields, respectively,
and $\Po$ denotes the path-ordering.
The loop $C$ is defined by ten coordinates, $x^\mu(s)$ and $y^i(s)$:
$x^\mu$ are the coordinates of the four-dimensional spacetime
in which the gauge theory lives, and this Wilson loop depends also on
the additional six ``coordinates'' $y^i$.
The corresponding object to this Wilson loop in the string theory side
is the string worldsheet whose boundary is specified by the loop
$C$. Then the conjectured relation
\cite{Rey:1998ik,Maldacena:1998im,Drukker:1999zq,Drukker:1999gy,
Polyakov:2000ti,Polyakov:2000jg,Semenoff:2002kk}
is
\begin{equation}
\exp\left(-A_\textrm{worldsheet}\right)
=\VEV{W(C)}_\textrm{SYM},
\label{e^A=W}
\end{equation}
where $A_\textrm{worldsheet}$ is the area of the classical solution of
the string worldsheet, and $\VEV{W(C)}_\textrm{SYM}$
is the expectation value of the Wilson loop operator.

On the other hand, there is another interesting correspondence
between the Wilson loop operator and the fundamental string in the
context of the string/gauge duality; the correspondence between the
Wilson loop operator and the string field.
In \cite{Fukuma:1997en} the loop equation of the Wilson loop operator
in type IIB matrix model was investigated and they argued that
the lightcone Hamiltonian of the string field can be derived from the loop
equation.
If there is a similar correspondence between the Wilson loop
operator $W(C)$ in four-dimensional ${\cal N} = 4$ SYM and the string
field $\Psi [X(s)]$,  it is natural to expect that the string field
lives in
a curved geometry, i.e., in ${\rm AdS}_5 \times {\rm S}^5$.
Hence the loop equation of $W(C)$ would have the same information as
the Hamiltonian of the string field on ${\rm AdS}_5 \times {\rm S}^5$.
Although the construction of string field theory (SFT) on
${\rm AdS}_5 \times {\rm S}^5$ spacetime is still a challenge,
there have been lots of developments in understanding the
SFT on the pp-wave background
\cite{Spradlin:2002ar,Spradlin:2002rv,Chu:2002eu,Pankiewicz:2002gs,
Pankiewicz:2002tg,Chu:2002wj,DiVecchia:2003yp,Pankiewicz:2003kj,
Pankiewicz:2003ap} 
which is obtained by taking the
Penrose limit of ${\rm AdS}_5 \times {\rm S}^5$ geometry.
Connections between the pp-wave string states and
the local operators in the gauge theory, i.e.,
the BMN operators, are also studied intensively
\cite{Berenstein:2002jq} (see also
\cite{Sadri:2003pr,Plefka:2003nb,Pankiewicz:2003pg} and references
therein).
Recently one of the present authors have shown that these BMN
operators emerge in the expansion of the Wilson loop operator with
respect to the fluctuations
$\delta C=\{\delta x^\mu ,\, \delta y^i \}$
of the loop $C$ \cite{Miwa:2005qz}:
\begin{align}
W(C) \sim
&\sum_J\frac{1}{J!}
\biggl\{
{\cal O}_{\textrm{ground}}^J
+
\delta x^\mu_0   {\cal O}_{\mu,0}^J
+
\delta \dot y^p_0 {\cal O}_{4+p,0}^J \notag \\
&+
\sum_n\delta x^\mu_{-n} \delta x^\nu_n {\cal O}_{\mu \nu ,n}^{J-1}
+
\sum_n\delta \dot y^p_{-n} \delta x^\mu_n {\cal O}_{4+p\,\mu ,n}^{J-1}
+
\sum_n\delta \dot y^p_{-n} \delta \dot y^q_n
{\cal O}_{4+p\,4+q ,n}^{J-1}
+ \cdots\biggr\},
\end{align}
where ${\cal O}_{\textrm{ground}}^J$, ${\cal O}_{M,0}^J$
and ${\cal O}_{M\,N,n}^J$ are the BMN operators
and the indices $p$ and $q$ run from $1$ to $4$.\footnote{
Note that we have expanded the Wilson loop operator with respect to the
fluctuations of the four coordinates $\delta x^\mu$
and the four ``winding number density''
$\delta \dot y^p$.
}
It is quite interesting that this
expansion resembles the expansion of a string field with respect
to the string states:
\begin{equation}
\Psi[X(s)] = \sum_A\braket{\delta X}{A}\, \psi_A(X_0) ,
\end{equation}
where $\{\ket{A}\}$ is a complete set of first-quantized string
states, and $\psi_A(X_0) $, which is a function of the center-of-mass
coordinate $X_0$ of the string, is the local field corresponding to
the string state $\ket{A}$.

Hence it is a very interesting theme to investigate the loop equation of
the Wilson loop operator in four-dimensional $\N = 4$ SYM
with the expectation that it would have the same information as the
equation of motion of the string field on the
${\rm AdS}_5\times{\rm S}^5$ background and on the pp-wave background
as well.
However, the loop equation in $\N=4$ SYM including its fermionic part
has not been completely established.\footnote{
Loop equations in four-dimensional $\N = 4$ SYM were studied
to check the correspondence \eqref{e^A=W}
\cite{Drukker:1999zq,Drukker:1999gy,Polyakov:2000ti,Polyakov:2000jg}.
In \cite{Drukker:1999zq,Drukker:1999gy}, they studied a special class
of loops satisfying the condition $\dot x^2 +\dot y^2=0$ or its
fermionic extension. In \cite{Polyakov:2000ti,Polyakov:2000jg},
they considered the
simple Wilson loop \eqref{bosonW} and argued that the contribution of
the scalars and gauginos is irrelevant for their analysis.
A manifestly supersymmetric formulation of the loop equation in
$\N =1$ SYM is given in \cite{Itoyama:1996ji,Itoyama:1997fc}.
}
The purpose of this paper is to construct the loop equation in
four-dimensional $\N = 4$ SYM as a first step of such an
investigation. We will also carry out a (partial) analysis of the
loop equation toward the identification of the Wilson loop
operator with the string field.

For explaining the problems in constructing the loop equation in $\N=4$
SYM, let us recapitulate the derivation of the loop equation in
bosonic Yang-Mills theory.
In bosonic Yang-Mills theory, the Wilson loop operator is given simply
by
\begin{equation}
W(C)=\Tr\Po\exp\left(i\int_0^l ds A_\mu(x(s))\dot x^\mu(s)\right)
\equiv \Tr\W_0^l(C).
\label{bosonW}
\end{equation}
Here, $\W_{u_1}^{u_2}(C)$ expresses the Wilson line defined on
the portion of the loop $C$ with the parameter region $[u_1,u_2]$.
In the rest of this paper we often omit the argument $C$ of
$\W_{u_1}^{u_2} (C)$ when it causes no confusion.
The starting point of deriving the loop equation is the following
formula for the functional derivative acting on
$\W_{u_1}^{u_2}$:
\begin{align}
\fDrv{}{x^\mu(s)}\W_{u_1}^{u_2}&=i\int_{u_1}^{u_2}\! du\,
\W_{u_1}^u\left[\p_\mu A_\nu(x(u))\,\dot x^\nu(u)\delta(u-s)
+A_\mu(x(u))\dot\delta(u-s)\right]\W_u^{u_2}
\notag\\
&=\W_{u_1}^s i\left(F_{\mu\nu}\dot x^\nu\right)_s\W_s^{u_2}
+\W_{u_1}^{u_2}i (A_\mu)_{u_2}\delta(s-u_2)
-i(A_\mu)_{u_1}\W_{u_1}^{u_2}\delta(s-u_1) ,
\label{d/dx^muW}
\end{align}
where $(F_{\mu\nu} \dot x^\nu)_{s}$, for example, is the
abbreviation of $F_{\mu\nu}(x(s))\dot x^\nu(s)$.
Using this formula twice we get
\begin{align}
&\frac{\delta}{\delta x^\mu(s_2)}
\frac{\delta}{\delta x_\mu(s_1)}W(C)
=\frac{\delta}{\delta x^\mu(s_2)}
\Tr\left[i (F^\mu{}_\nu \dot x^\nu)_{s_1} \W_{s_1}^{s_1 + l}
\right]
\notag\\
&=
  \Tr
  \left[
    i \bigl( F^\mu{}_\nu \dot x^\nu \bigr)_{s_1}
    \W_{s_1}^{s_2}
    i \bigl( F_{\mu \rho} \dot x^\rho \bigr)_{s_2}
    \W_{s_2}^{s_1+l}
  \right]
  +
  \delta(s_1-s_2)
  \Tr
  \left[
    i \bigl( D_\mu F^\mu{}_\nu  \dot x^\nu \bigr)_{s_1}
    \W_{s_1}^{s_1+l}
  \right].
  \label{d^2W/dxdx}
\end{align}
We call the first term in the final form of \eqref{d^2W/dxdx}
\nd-term and the second term $\delta$-term.
It is important that the $\delta$-term is proportional to the LHS of the
equation of motion (EOM), $D_\mu F^\mu{}_\nu=0$.
Let us consider the expectation value of \eqref{d^2W/dxdx} or that of
the product of \eqref{d^2W/dxdx} and other Wilson loop operators.
Then the $\delta$-term can be evaluated as follows:
\begin{align}
   \int \!\! \D A_\mu
     &\Tr\!\left[
     it^a (D_\mu F^{\mu\nu} \dot x_\nu)^a_{s_1}
     \W_{s_1}^{s_1 + l}
   \right]
   (\cdots)
   e^{iS}
   =
   g^2 \int \! \! \D A_\mu
   \Tr\!
   \left[
    t^a
    \left(
    \dot x_\nu
     \frac{\delta}{\delta A_\nu^a} e^{iS}
    \right)_{\! \! s_1} \hspace{-0.3cm}
    \W_{s_1}^{s_1+l}
   \right]
   (\cdots)
\notag \\
&= -g^2\int \! \! \D A_\mu
  \Tr\!
  \left[
    t^a \dot x_\nu (s_1)
      \frac{\delta}{\delta A_\nu^a(x(s_1))}
      \W_{s_1}^{s_1+l}
  \right]
  (\cdots)
  e^{iS},
  \label{intbyparts}
\end{align}
where $t^a$ $(a= 1,\cdots,N^2-1)$ are the generators of the SU($N$)
gauge group, dots ($\cdots$)
express the possible other Wilson loop operators,
and $g$ is the Yang-Mills coupling constant.
We have performed functional integration by parts in obtaining the
final expression.\footnote{
If some operators in $(\cdots)$ lie on the loop $C$,
there are other contributions to the RHS of \eqref{intbyparts} which
arise when the functional derivative acts on such operators. Here we
just neglect such situations.
}
The functional derivative with respect to $A^a_\nu(x(s_1))$
divides the Wilson loop into two parts
%$[s_1, s]$ and $[s , s_1+l]$,
and we have, in functional integration,
\begin{align}
  \frac{\delta}{\delta x^\mu(s_2)}
  \frac{\delta}{\delta x_\mu(s_1)}
  W(C)&=
  \Tr\left[
    i \bigl( F^\mu{}_\nu \dot x^\nu \bigr)_{s_1}
    \W_{s_1}^{s_2}
    i \bigl( F_{\mu \rho} \dot x^\rho \bigr)_{s_2}
    \W_{s_2}^{s_1+l}
  \right] \notag \\
  & \hspace{-1cm}
  - i \frac{g^2}{2}
  \delta(s_1 - s_2) \int_{s_1}^{s_1+l}\!\!
  ds\,\delta^{(4)}\bigl(x(s)-x(s_1)\bigr)\dot x^\nu(s_1)\dot x_\nu(s)
  W(C_1) W(C_2)
  , \label{WW-W/N}
\end{align}
where the loop $C_1$ ($C_2$) is the part of the loop $C$
with the parameter region [$s_1,s$] ([$s,s_1+l$]).\footnote{
In this step we use the following formulas
$$
\fDrv{}{A_\mu^a(x(s))}\W_{u_1}^{u_2}=
\int_{u_1}^{u_2}\! du\,\W_{u_1}^u it^a\dot x^\mu(u)
\,\delta^{(4)}\bigl(x(u)-x(s)\bigr)\W_u^{u_2} ,
$$
and
$$
(t^a)_{ij} (t^a)_{kl}
  =
  \frac{1}{2}
  \left(
    \delta_{il} \delta_{jk}
    -
    \frac{1}{N} \delta_{ij} \delta_{kl}
  \right).
$$
In \eqref{WW-W/N} and throughout this paper, we neglect the $1/N$ term
in the second formula.
}
We call \eqref{d^2W/dxdx} and \eqref{WW-W/N} ``loop equation'' in this
paper.
The loop equation in bosonic Yang-Mills theory has been used to study
the area-law property of the Wilson loop (see
\cite{Polyakov:1987ez,Makeenko:2002uj} and references therein).

We would like to extend the above derivation of the loop equation to
the four-dimensional $\N=4$ SYM. Concretely, we have to give the SYM
extension of both the Wilson loop operator and the quadratic
functional derivative with respect to the loop
coordinates in such a way that the $\delta$-term which is multiplied by
$\delta(s_1-s_2)$ is proportional to the EOM and hence it gives the
non-linear term in the Wilson loop.
Our Wilson loop operator in $\N=4$ SYM is given by modifying
\eqref{intsuperW} to include the gaugino fields.
Accordingly, the loop $C$ is specified by two fermionic spinor
coordinates, $\zeta(s)$ and $\eta(s)$, as well as $4+6$ bosonic
coordinates, $x^\mu(s)$ and $y^i(s)$.
The coordinate $\zeta(s)$ has already appeared in the literature
\cite{Drukker:1999zq,Drukker:1999gy}. Its mass-dimension is $-1/2$ and
it has the same chirality as the gaugino.
On the other hand, another fermionic coordinate $\eta(s)$ has
mass-dimension $-2/3$ and the opposite chirality to that of $\zeta(s)$
and gaugino. Therefore, we can consider the quadratic functional
derivative
$(\delta/\delta\eta(s_2))(\delta/\delta\ol{\zeta}(s_1))$
which has the same mass-dimension $2$ as
$(\delta/\delta x^\mu(s_2))(\delta/\delta x_\mu(s_1))$
and $(\delta/\delta y^i(s_2))(\delta/\delta y^i(s_1))$.
By taking as the total quadratic functional derivative for the loop
equation a suitable linear combination of the above three, we have
succeeded in determining the dependence of $W(C)$ on the fermionic
coordinates so that the $\delta$-term may vanish to quadratic order in
$\zeta$ and $\eta$ if we use the EOM of $\N=4$ SYM.

In this way we can obtain the $\N=4$ SYM version of the loop equation
\eqref{WW-W/N}. For our application of the loop equation to the
analysis of the AdS/CFT correspondence, in particular, the
identification of the Wilson loop operator as the string field on
${\rm AdS}_5 \times{\rm  S}^5$, we have to consider the coincident limit
$s_1\to s_2$ of the quadratic functional derivative.
This limit is singular and needs some kind of regularization.
In this paper, we adopt the regularization of replacing the massless
free propagator $1/x^2$ by $1/(x^2+\epsilon^2)$, and extract the most
singular part of order $1/\epsilon^4$ in the loop equation.
We find that the resulting equation for the Wilson loop resembles
the Hamiltonian constraint of bosonic string
on ${\rm AdS}_5\times{\rm S}^5$ if we identify the UV regularization
parameter $\epsilon$ with the radial coordinate of ${\rm AdS}_5$.

The rest of this paper is organized as follows.
In section \ref{Wilson44} and \ref{0thLeq}, we consider the loop
equation in $\N=4$ SYM at the lowest order in the fermionic
coordinates: We derive the loop equation in section \ref{Wilson44},
and then in section \ref{0thLeq} we pick up the most singular and
linear part of the loop equation and compare it with the Hamiltonian
constraint of the bosonic string on the
${\rm AdS}_5 \times {\rm S}^5$.
In section \ref{General} we extend our loop equation to the quadratic
order in the fermionic coordinates.
Section \ref{Conclusion} is devoted to the conclusion and
discussions. Our notations and conventions are summarized in
appendix \ref{notation}.
Details of the calculations used in section \ref{General} are given in
appendix \ref{calc}.
In appendix \ref{free} we calculate the most singular and linear part
of the loop equation to quadratic order in fermionic coordinates.
In appendix \ref{generaldiff} we consider more
general functional derivatives than those we consider in
section \ref{0thLeq} and \ref{General}.

\section{Loop equation in $\bm{\N=4}$ SYM I:
the lowest order in $\bm{\zeta}$ and $\bm{\eta}$}
\label{Wilson44}

As we explained in the previous section, the loop equation in
four-dimensional $\N = 4$ SYM depends on the fermionic loop coordinates
$\zeta(s)$ and $\eta(s)$ as well as the bosonic coordinates $x^\mu(s)$
and $y^i(s)$. In this section, we will derive the loop equation at the
lowest order in $\zeta(s)$ and $\eta(s)$; namely, we consider the loop
equation by putting $\zeta(s)=\eta(s)=0$ from outside (i.e., after
functional differentiations with respect to the loop coordinates).
Extension to quadratic order in $\zeta(s)$ and $\eta(s)$
is given in section \ref{General}.

First, our notations for the four-dimensional $\N = 4$ SYM are as
follows. The field content of this theory is
one gauge field, six scalar fields and four gauginos
which are four dimensional Weyl spinors. In this paper we adopt the
ten-dimensional $\N = 1$ notation: the gauge field and the scalar
fields are expressed
by $A_\mu$ ($\mu = 0,\ldots,3$) and $A_{3+i}$ ($i=1,\ldots,6$),
respectively, and we combine four Weyl spinors to make one
ten-dimensional  Majorana-Weyl spinor $\Psi$.
We have summarized our notations in appendix \ref{notation}.
Our action of the four-dimensional ${\cal N} = 4$ SYM is
\begin{align}
  {\cal L}
  =
  -\frac{1}{ g^2}
  \Tr\left(
    \frac{1}{2}
      F_{MN} F^{MN}
    +
    i
      \overline\Psi
      \Gamma^M D_M
      \Psi
  \right),
\label{action}
\end{align}
with $M$ and $N$ running from 0 to 9.
We have defined the field strengths and the covariant derivatives as
follows:
\begin{align}
  &F_{\mu \nu}
  =
  \partial_\mu A_\nu - \partial_\nu A_\mu
  +
  i
  \left[
    A_\mu , A_\nu
  \right],\quad
  F_{\mu , 3+i} = - F_{3+i , \mu}
  =
  D_\mu A_{3+i}, \quad
  F_{3+i , 3+j}  = i \left[ A_{3+i} , A_{3+j} \right], \notag \\
  &D_\mu \Op=
  \partial_\mu \Op + i \left[ A_\mu , \Op \right], \quad
  D_{3+i} \Op = i \left[ A_{3+i} , \Op \right].
\end{align}
The ten-dimensional Dirac matrices $\Gamma_M$ satisfy the following
Clifford algebra:
\begin{align}
  \left\{ \Gamma_M , \Gamma_N \right\} = 2 \eta_{MN},
  \label{Clifford}
\end{align}
where $\eta_{MN} = {\rm diag} (-1,1,\cdots,1)$ is the
ten-dimensional flat metric.
We will also use the four-dimensional flat metric
$\eta^{(4)}_{MN} = {\rm diag}(-1,1,1,1,0,\cdots,0)$.
The action \eqref{action} is invariant under the SUSY transformation
$\delta_\zeta$:
\begin{align}
  \delta_\zeta A_M = - i \ol{\zeta} \Gamma_M \Psi, \quad
  \delta_\zeta \Psi= \frac{1}{2} F_{MN} \Gamma^{MN} \zeta ,
  \label{SUSY}
\end{align}
with $\Gamma^{MN}=(1/2)\bigl[\Gamma^M,\Gamma^N\bigr]$.
In \eqref{SUSY}, $\zeta$ is the fermionic variable with the same
chirality as that of $\Psi$.

Let us start constructing the loop equation. As we stated in the
previous section, we have to give the $\N=4$ SYM extension of both the
Wilson loop operator and the quadratic functional derivative
with respect to the loop coordinates. The guiding principle of
this extension is that the $\delta$-term, namely the term multiplied by
$\delta(s_1-s_2)$, be proportional to the EOM of $\N=4$ SYM (see
\eqref{d^2W/dxdx}).
The EOM of four-dimensional ${\cal N} = 4$ SYM is given by
\begin{align}
  &D_M F^M{}_N - \ol{\Psi} \Gamma_N \Psi = 0, \label{bosoneq}\\
  &\DS \Psi = 0, \label{fermioneq}
\end{align}
with $\DS = \Gamma^M D_M$.
Note that \eqref{bosoneq} expresses the EOM of the scalar fields
as well as the gauge field.
The first term of \eqref{bosoneq} can be obtained by
considering the Wilson loop operator \eqref{intsuperW}:
\begin{align}
  W(C) = \Tr \Po
    \exp
    \left(
      i \int_0^l ds A_M(x(s)) \dot X^M (s)
    \right)
  \equiv
  \Tr \W_0^l , \label{superW}
\end{align}
where we have introduced the ten-dimensional loop coordinates $X^M$
with $X^\mu=x^\mu$ and $X^{3+i}=y^i$.
By performing the ten-dimensional functional differentiation and
repeating the derivation of \eqref{d^2W/dxdx} we obtain
\begin{align}
  &\frac{\delta}{\delta X_M(s_2)}
  \frac{\delta}{\delta X^M(s_1)}
  W(C) \notag \\
  &=
  \Tr
  \left[ \!
    \bigl(i
    F^M{}_N \dot X^N
    \bigr)_{\! s_1}\!
    \W_{s_1}^{s_2}
    \bigl( i
    F_{M P} \dot X^P
    \bigr)_{\!s_2}\!
    \W_{s_2}^{s_1+l}
  \right]
  +
  \delta(s_1-s_2)
  \Tr
  \left[ \!
     \bigl( i D_M F^M{}_N \dot X^N\bigr)_{\!s_1}
    \!\W_{s_1}^{s_1+l}
  \right]. \label{loop_eq_AS}
\end{align}
We see that the $\delta$-term of \eqref{loop_eq_AS} contains correctly
the first term of \eqref{bosoneq}.
It is obvious that in order to reproduce completely the LHS of
\eqref{bosoneq} including the gaugino current, we have to introduce
the fermionic fields in the Wilson loop.
Previously the following fermionic extension of the Wilson loop
operator obtained as the ``SUSY transformation'' of \eqref{superW} has
been considered \cite{Drukker:1999zq,Drukker:1999gy}:
\begin{align}
  W(C) = \Tr \Po
    \exp
    \left(
      i
      \int_0^l ds
      \A_M (x(s),\zeta(s)) \dot X^M (s)
    \right),
\label{prevsupersuperW}
\end{align}
where $\A_M$ is the finite SUSY transformation of $A_M$:
\begin{align}
  \A_M(x,\zeta)
  &=
  A_M
  +
  \delta_{\zeta} A_M
  +
  \frac{1}{2}
  \delta_{\zeta}^2 A_M
  +\cdots \notag \\
  &=
  A_M
  -
  i \ol{\zeta} \Gamma_M \Psi
  -
  \frac{i}{4} F_{NP} \ol{\zeta} \Gamma_M \Gamma^{NP} \zeta
  + \cdots.
  \label{superA}
\end{align}
In \eqref{prevsupersuperW}, the parameter $\zeta$ is promoted to a
$s$-dependent fermionic loop coordinate $\zeta(s)$.
However, it seems hard to reproduce the complete EOM of
\eqref{bosoneq} by adopting this type of Wilson loop and a
simple quadratic functional derivative.

In this paper we consider another type of Wilson loop operator by
introducing an additional fermionic coordinate $\eta$.
Our motivation of introducing such a coordinate is the dimension of
the functional differential operator.
The quadratic functional derivative on the LHS of
\eqref{loop_eq_AS} has mass-dimension 2, and we want a differential
operator with respect to fermionic coordinates whose mass-dimension
is also 2.
Because the mass-dimension of $\zeta$ is $-1/2$, we are lead to the
idea of introducing an additional fermionic variable $\eta$ which
carries mass-dimension $-3/2$ and therefore allows us to consider the
following differential operator with mass-dimension 2:
\begin{align}
  \frac{\delta}{\delta \eta_\alpha(s_2)}
  \frac{\delta}{\delta \overline\zeta_\alpha(s_1)},
\label{d^2/dedz}
\end{align}
where the $\alpha$ is the spinor index.
The chirality of $\eta$ must be opposite to that of $\zeta$ and
$\Psi$.

Next we must fix the dependence of the Wilson loop operator on these
two fermionic coordinates $\zeta$ and $ \eta$.
We have already given a well motivated way of introducing the
coordinate $\zeta$, i.e., through SUSY transformation
\eqref{superA}.\footnote{
In section \ref{General} we will find that the coefficients of
the terms $\delta_\zeta^n A_M$ in \eqref{superA} need to be modified
for $n\ge 2$. Here we need only the terms with $n=0$ and $1$.
}
On the other hand, we do not know such an origin of the variable
$\eta$.
In any case, the dependence of the Wilson loop on $\eta$ should be
determined from the requirement that the functional derivative
\eqref{d^2/dedz} acting on the Wilson loop supply the needed gaugino
current term in \eqref{bosoneq}.
Actually, if we set $\zeta = \eta = 0$ from outside,
this requirement can be fulfilled by considering following operator:
\begin{align}
  W(C) =
  \Tr \Po
    \exp
    \left(
      i
      \int_0^l ds
      \Bigl(
        A_M (x(s))\dot X^M(s)
        -
        i \ol{\zeta}(s) \Gamma_M \Psi(x(s)) \dot X^M(s)
        +
        \overline\Psi (x(s)) \dot \eta (s)
      \Bigr)
    \right).
\label{lowestW}
\end{align}
Let us consider $K_{\beta_1}W(C)\bigr|_{\zeta=\eta=0}$ with quadratic
functional derivative $K_{\beta_1}$ defined by
\begin{align}
K_{\beta_1}=
    \frac{\delta}{\delta X_M (s_2)}
    \frac{\delta}{\delta X^M (s_1)}
    +
    \beta_1
    \frac{\delta}{\delta \eta (s_2)}
    \frac{\delta}{\delta \ol{\zeta} (s_1)} ,
\label{K_beta1}
\end{align}
and $W(C)$ given by \eqref{lowestW}.
In \eqref{K_beta1}, $\beta_1$ is a numerical
coefficient to be determined below.
Similarly to \eqref{d/dx^muW}, the first derivatives of the
Wilson line of \eqref{lowestW} are given by
\begin{align}
  \frac{\delta }{\delta X^M(s)} \W_{u_1}^{u_2}
  &=
  \W_{u_1}^s
  \bigl(
    i F_{MN} \dot X^N
    +
    \ol{\zeta} \Gamma_{[N} D_{M]} \Psi \dot X^N
    +
    i D_M \ol{\Psi} \dot \eta
    -
    \dot {\ol{\zeta}} \Gamma_M \Psi
  \bigr)_s
  \W_s^{u_2} \notag \\
  &\quad+
  \W_{u_1}^s
  \bigl[
    \ol{\zeta} \Gamma_M \Psi
    ,
    \ol{\zeta} \Gamma_N \Psi \dot X^N
    +
    i \ol{\Psi} \dot \eta
  \bigr]_s
  \W_s^{u_2} \notag \\
  &\quad
  +
  \W_{u_1}^{u_2}
  \left( i A_M + \ol{\zeta} \Gamma_M \Psi \right)_{u_2}
  \delta(s-u_2)
  -
  \left( i A_M + \ol{\zeta} \Gamma_M \Psi \right)_{u_1}\W_{u_1}^{u_2}
  \delta(s-u_1 ),  \\
  \frac{\delta}{\delta \ol{\zeta}(s)}
  \W_{u_1}^{u_2}
  &=
  \W_{u_1}^s
  \bigl(\Gamma_M \Psi \dot X^M\bigr)_s
  \W_s^{u_2},
  \\
  \frac{\delta}{\delta \eta(s)}
  \W_{u_1}^{u_2}
  &=
  \W_{u_1}^s
  \bigl(
    i D_M \ol{\Psi} \dot X^M
    +
    i
    \bigl[
      \ol{\zeta} \Gamma_M \Psi \dot X^M + i \ol{\Psi} \dot \eta
      ,
      \ol{\Psi}
    \bigr]
  \bigr)_s
  \W_s^{u_2} \notag \\
  & \qquad
  + \W_{u_1}^{u_2}(-i \ol{\Psi})_{u_2}\delta(s-u_2)
  - (-i \ol{\Psi})_{u_1}\W_{u_1}^{u_2}\delta (s-u_1).
\end{align}
Using these formulas, we obtain
\begin{align}
K_{\beta_1}W(C)\Bigr|_{\zeta = \eta = 0}&=
  \Tr \Po
  \left[
    \left(
      (iF_{MN} \dot X^N )_{s_1}  (i F^M{}_P \dot X^P)_{s_2}
      -
      \beta_1
    \bigl( \Gamma_N \Psi \dot X^N \bigr)_{s_1}
    \bigl(i D_P \ol{\Psi} \dot X^P \bigr)_{s_2}
  \right)
    \W_0^l
  \right] \notag \\
  &\qquad +
  \delta(s_1 - s_2)
    \Tr
  \left[
    i
    \Bigl(\bigl(
      D_M F^M{}_N
      -2\beta_1
      \ol{\Psi}  \Gamma_N \Psi
      \bigr) \dot X^N
    \Bigr)_{s_1}
      \W_{s_1}^{s_1+l}
  \right]
 . \label{pre_loopeq1}
\end{align}
We find that the last term of \eqref{pre_loopeq1} contains the LHS
of the EOM \eqref{bosoneq} if we set
\begin{equation}
\beta_1=\frac12 .
\label{beta1=1/2}
\end{equation}
Then carrying out the functional integration by parts as we did in
\eqref{intbyparts}, we get
\begin{align}
&K_{\beta_1=\frac12}W(C)\Bigr|_{\zeta=\eta=0}=
  \dot X^N (s_1) \dot X^P(s_2)
  \Tr \Po \!
  \left[
    \left(
      (i F_{MN})_{s_1} (i F^M{}_P)_{s_2}
      -
      \frac{1}{2}
    \left(\Gamma_N \Psi \right)_{s_1}
    \left(i D_P \ol{\Psi} \right)_{s_2}
  \right)
    \W_0^l
  \right] \notag \\
&\qquad\qquad\qquad
 - i \delta(s_1-s_2) \frac{g^2}{2}
  \int_{s_1}^{s_1+l}\! ds\,
  \delta^{(4)}\bigl(x(s)-x(s_1)\bigr)
    \dot X_N(s) \dot X^N(s_1)
      W(C_1) W(C_2)
. \label{pre_loopeq2}
\end{align}
In this way we have derived the loop equation by setting
$\zeta = \eta = 0$ from outside.

Before closing this section, we will make some comments on the last
$\delta$-term of \eqref{pre_loopeq2}.
Recall that $C_1$ ($C_2$) is the part of the loop $C$ with its
parameter region [$s_1,s$] ([$s,s_1 + l$]).
The existence of four-dimensional delta function
$\delta^{(4)}( x(s) - x(s_1) )$ implies that the integration with
respect to $s$ has contributions only from points satisfying
$x(s)=x(s_1)$.
There are two types of such contributions.
One is the contribution from the point $s=s_1$, and this exists for
any loop $C$. The other kind of contribution arises if the loop has
self-intersecting points and if $x^\mu(s_1)$ is just one of these
points.
For the former contribution, either of the two loops $C_1$ and $C_2$
becomes trivial and we have
\begin{align}
  W(C_1) W(C_2) = \Tr[1] W(C) = N W(C).
\end{align}
For the latter contribution, none of the two loops become trivial and
we should regard this term as the interacting part of the loop
equation.

\section{Loop equation and Hamiltonian constraint}
\label{0thLeq}

In this section we will consider the limit $s_1 \to s_2$ in
the loop equation \eqref{pre_loopeq2}.
In this limit, \eqref{pre_loopeq2} has some singularities.
We will extract the most singular contribution to the terms linear in
the Wilson loop. Namely, we neglect the contribution from the
self-intersecting points of $C$ in the last term of
\eqref{pre_loopeq2}.
We compare the resulting linear equation for the Wilson loop with the
Hamiltonian constraint of bosonic string on
${\rm AdS}_5\times{\rm S}^5$.
Recall that we call the first and the second term on the RHS of
\eqref{pre_loopeq2} \nd-term and $\delta$-term, respectively.

\subsection{Linear and the most singular part of the loop
  equation}
\label{LMS}

First we will consider the singular part of the \nd-term
of \eqref{pre_loopeq2} in the limit $s_1 \to s_2$.
Singularities arise when two operators at $s_1$ and $s_2$ collide with
each other. We evaluate these singularities by taking the
contraction of the two operators by using the following UV regularized
free propagators:
\begin{align}
  \wick{1}
  {
    <1 A_M^a(x) >1A_N^b
  }(\tilde x)
  =
  \frac{g^2}{4 \pi^2}
  \frac{\delta^{ab} \eta_{MN}}{(x-\tilde x)^2 + \epsilon^2},
  \qquad
  \wick{1}
  {
    <1 \Psi^a_\alpha (x)  >1 {\overline{\Psi}}^b_\beta
  }(\tilde x)
  =
  \frac{i g^2}{4 \pi^2}
  \slashchar{\partial}_{\alpha \beta}
  \frac{\delta^{ab}}{(x-\tilde x)^2 + \epsilon^2},
\label{pro}
\end{align}
where $\epsilon$ is the short distance cutoff parameter.
Here, we consider only the leading order terms in the SYM coupling
constant $g$. Using \eqref{pro} we have
\begin{align}
  \wick{1}
  {
    \partial_M <1 A_N^a(x)
    \partial_P >1 A_Q^b
  }(\tilde x)\Bigr|_{x=\tilde x}
 =
  \frac{g^2}{4 \pi^2}
  \frac{2\delta^{ab} \eta^{(4)}_{MP}\eta_{NQ}}{\epsilon^4},
  \quad
  \wick{1}
  {
    \partial_M
    <1\Psi^a_\alpha(x)
    >1 {\overline \Psi}^b_\beta
  }
  (\tilde x)\Bigl|_{x=\tilde x}
  =
  \frac{i g^2}{4 \pi^2}
  \frac{-2 (\Gamma^N)_{\alpha \beta} \eta^{(4)}_{MN}
  \delta^{ab}}
  {\epsilon^4}.
  \label{DPP}
\end{align}
Using the first equation of \eqref{DPP}
and to the leading order in $g$, we have the following contraction
of two field strengths:
\begin{align}
  \wick{1}
  {
    <1 F^a_{MN}(x) >1 F^b_{PQ}
  }(\tilde x)\Bigr|_{x=\tilde x}
  =
  \frac{ 2 g^2 \delta^{ab}}{4 \pi^2 \epsilon^4}
  \left(
    \eta^{(4)}_{MP}
    \eta_{NQ}
    -
    \eta^{(4)}_{MQ}
    \eta_{NP}
    -
    \eta^{(4)}_{NP}
    \eta_{MQ}
    +
    \eta^{(4)}_{NQ}
    \eta_{MP}
  \right)
  . \label{FF}
\end{align}
{}From these rules the singular part of
the \nd-term of \eqref{pre_loopeq2} can be evaluated to the leading
order in $g$ as
\begin{align}
  \lim_{s_1 \to s_2}
  \biggl\{
  \wick{1}
  {
    i (<1 F_{MN} )_{s_1}
    i ( >1 {F}^M{}_P )_{s_2}
  }
  -
  \frac{1}{2}
  \wick{1}
  {
    ( \Gamma_N <1 \Psi )_{s_1}
    ( i D_P >1 {\overline \Psi} )_{s_2}
  }
  \biggr\}
  &=
  \frac{\lambda}{\pi^2}
  \frac{1}{\epsilon^4}
  \left(
    -
    \left(
      \eta_{NP}
      +
      2 \eta^{(4)}_{NP}
    \right)
    +
    2 \eta^{(4)}_{NP}
  \right) \notag \\
  &=
  -
  \frac{\lambda}{ \pi^2}
  \frac{ \eta_{NP}}{\epsilon^4},
\label{FF+PsiPsi}
\end{align}
where $\lambda=g^2 N$ is the 't\,Hooft coupling.
It is interesting to observe the following: The contribution to the
singular part from the bosonic fields and that from the fermionic
fields do not have ten-dimensional covariance separately.
However, once they are added using the coefficient $\beta_1=1/2$
\eqref{beta1=1/2}, we regain the ten-dimensional covariance as in
\eqref{FF+PsiPsi}.
Using \eqref{FF+PsiPsi}, the most singular part of the \nd-term in
\eqref{pre_loopeq2} is given as follows:
\begin{align}
  \mbox{\nd-term of \eqref{pre_loopeq2}}
  =
  -
  \frac{\lambda}{\pi^2}
  \frac{\dot X^M (s_1) \dot X_M (s_1)}{\epsilon^4}
  W(C)
  +O(1/\epsilon^{3}) ,
  \label{bosonnondeltafree}
\end{align}
to the leading order in the coupling constant.

Next let us turn to the linear part in $W(C)$ of the $\delta$-term of
\eqref{pre_loopeq2}.
We already explained that the linear part in $W(C)$ comes from the
region $s \sim s_1$ in the $s$-integration of \eqref{pre_loopeq2}.
Note that there are two kinds of singularities contained in the last
term of \eqref{pre_loopeq2} with $s_1=s_2$.  One is $\delta(s_1-s_1)$
multiplying \eqref{pre_loopeq2}. Besides this, the $s$-integration
around $s=s_1$ is divergent without putting $s_1=s_2$.
We will treat the former singularity in the next subsection.
For the latter singularity, we adopt the following regularized
four-dimensional delta function:
\begin{align}
    \delta^{(4)} (x)
  &=
  \frac{2i}{\pi^2}
  \frac{\epsilon^2}{\left( x^2 + \epsilon^2 \right)^3}.
\label{regdelta}
\end{align}
This regularization is consistent with the propagators \eqref{pro}
in the sense that
$\p_\mu\p^\mu(x^2+\epsilon^2)^{-1}=i(2\pi)^2\delta^{(4)}(x)$.
Using this delta function, we can evaluate the contribution to the
$s$-integration from the region $s\sim s_1$ as follows:
\begin{align}
  \textrm{ $\delta$-term of \eqref{pre_loopeq2}}
  &\sim
  - i \delta(s_1 - s_1) \frac{\lambda}{2}
  \dot X_N (s_1) \dot X^N (s_1) W(C)
  \int\! ds\,
  \frac{2i}{\pi^2}
  \frac{\epsilon^2}
  {\left( (s - s_1)^2 (\dot x(s_1) )^2 + \epsilon^2    \right)^3},
  \notag\\
  &=
  \frac{3\lambda}{8 \pi}
  \frac{\dot X^N(s_1) \dot X_N (s_1)}{\epsilon^4} W(C)
  \times
    \frac{\epsilon\,\delta (s_1 -s_1) }{\sqrt{( \dot x (s_1))^2}}
  +
  O(1/\epsilon^3) .
  \label{bosondeltafree}
\end{align}
Here we have assumed that
$\epsilon\delta (s_1 -s_1)/\sqrt{( \dot x (s_1))^2}$ is a finite
quantity of order $\epsilon^0$ (see the next subsection).

Finally, from \eqref{bosonnondeltafree} and \eqref{bosondeltafree}, we
obtain the following expression for the loop equation
\eqref{pre_loopeq2} with $s_1=s_2$:
\begin{align}
  \left(
    -\frac{\delta}{\delta X_M(s_1)}
    \frac{\delta}{\delta X^M(s_1)}
    -
    \frac{1}{2}
    \frac{\delta}{\delta \eta(s_1)}
    \frac{\delta}{\delta \ol{\zeta} (s_1)}
    +
    \lambda \kappa
    \frac{\dot X^M(s_1) \dot X_M(s_1)}{\epsilon^4}
  \right)
  W(C)
  \Biggr|_{\zeta=\eta=0}
  \hspace{-0.5cm}
  +
  \ldots
  =
  0 , \label{0th_loopeq}
\end{align}
where $\kappa$ is defined by
\begin{align}
\kappa =\frac{3}{8\pi}
\frac{\epsilon\,\delta(s_1 - s_1)}{\sqrt{(\dot x(s_1))^2}}
-\frac{1}{\pi^2}.
\label{kappa}
\end{align}
In \eqref{0th_loopeq}, the dots $\ldots$ denote the less singular
terms in $\epsilon$, higher order terms in $g$, and the non-linear
terms in the Wilson loop.
In $\kappa$ of \eqref{kappa}, the first term is the contribution from
the $\delta$-term and the second term is from the \nd-term.

\subsection{Hamiltonian constraint on $\bm{{\rm AdS}_5 \times {\rm S}^5}$}

Let us compare \eqref{0th_loopeq} with the Hamiltonian constraint
of bosonic string on ${\rm AdS}_5 \times {\rm S}^5$.
The latter can be derived from the Polyakov action:
\begin{align}
  S_{\rm Polyakov}
  &=
  -
  \frac{1}{4 \pi \alpha'}
  \int d^2 \sigma
  \sqrt{-g}
  g^{ab} G_{MN}
  \partial_a {\cal X}^M(\sigma)
  \partial_b {\cal X}^N(\sigma)
  =
  \int d^2 \sigma {\cal L}_{\rm Polyakov},
\end{align}
where ${\cal X}^M(\sigma)$ ($M,N = 0,\ldots,9$) is the string
coordinate, $g_{ab}$ ($a,b = 0,1$) is the worldsheet metric and
$G_{MN}$ is the spacetime metric.
The Hamiltonian constraint is
\begin{align}
  (2 \pi \alpha')^2
  G^{MN} {\cal P}_M {\cal P}_N
  +
  G_{MN} \partial_1 {\cal X}^M \partial_1 {\cal X}^N
  =
  0.\label{Hamiltonian2}
\end{align}
We have introduced momentum ${\cal P}_M$ conjugate to ${\cal X}^M$:
\begin{align}
  {\cal P}_M (\sigma)
  =
  \frac{\partial {\cal L}_{\rm Polyakov}}
  {\partial( \partial_0 {\cal X}^M (\sigma))}
  =
  -
  \frac{1}{2 \pi \alpha'}
  \sqrt{-g}g^{0a}
  G_{MN} \partial_a {\cal X}^N(\sigma).
\end{align}
In the Poincare coordinate of ${\rm AdS}_5 \times {\rm S}^5$ geometry
with the line element
\begin{align}
  ds^2
  = G_{MN} d{\cal X}^M d{\cal X}^N
  = \frac{R^2}{Y^2}
  \left(
    dx^\mu dx_\mu + dY^i dY^i
  \right), \label{Poincare}
\end{align}
\eqref{Hamiltonian2} becomes
\begin{align}
  (2 \pi \alpha')^2
  \frac{Y^2}{R^2}
  (
  {\cal P}_\mu {\cal P}^\mu
  +
  {\cal P}^Y_i {\cal P}^Y_i
  )
  +
  \frac{R^2}{Y^2}( \dot x^\mu \dot x_\mu + \dot Y^i \dot Y^i )
  =0, \label{PoincareHamilton}
\end{align}
with ${\cal X}^M=(x^\mu,Y^i)$ and
${\cal P}_M=({\cal P}_\mu,{\cal P}^Y_i)$.
In \eqref{Poincare} and \eqref{PoincareHamilton}, the index $\mu$ is
raised or lowered using the four-dimensional flat metric
$\eta^{(4)}_{\mu\nu}$, and the dot denotes the derivative with respect to
$\sigma^1$.
Let us identify $x^\mu$ with the four-dimensional coordinate where
the SYM lives. The remaining six coordinates $Y^i$ in
\eqref{Poincare}, however, does not directly correspond to
$y^i =X^{3+i}$ of the Wilson loop operator \eqref{lowestW}.
Actually these two sets of coordinates $y^i$ and $Y^i$ should be
related through T-duality \cite{Drukker:1999zq,Fukuma:1997en}.
T-duality on curved backgrounds is a subtle matter, but
here we just assume that these two coordinates are related through
(here we ignore the ordering problem)
\begin{align}
  \dot y^i
  &
  =  2 \pi \alpha' G^{3+i,M} {\cal P}_M ,
  \quad
  - i \frac{\delta}{\delta y^i}
  =  \frac{1}{2 \pi \alpha'} G_{3+i,M} \dot{\cal X}^M,
  \label{T-dual1}
\end{align}
where we identify $\sigma_1$ with the parameter $s$ of the Wilson loop.
Using the momentum $P_M(s)$ conjugate to the loop coordinate
$X^M(s)$ defining the Wilson loop,
\begin{align}
  P_M(s) = - i\frac{\delta}{\delta X^M(s)},
\end{align}
the above relation \eqref{T-dual1} can be rewritten as
\begin{align}
  \dot y^i
  =  2 \pi \alpha' \frac{Y^2}{R^2} {\cal P}^Y_i
  ,\quad
  P_{3+i}
  = \frac{1}{2 \pi \alpha'} \frac{R^2}{Y^2} \dot Y^i.
\end{align}
Therefore, the Hamiltonian constraint \eqref{PoincareHamilton} is
expressed in terms of the coordinate $X^M =(x^\mu,y^i)$ and its
conjugate $P_M$ as
\begin{align}
  ( 2 \pi \alpha' )^2 \frac{Y^2}{R^2}
  P_M P^M
  +
  \frac{R^2}{Y^2}
  \dot X^M \dot X_M
  =
  0,
\label{HC}
\end{align}
where the index $M$ should be raised or lowered using the
ten-dimensional flat metric $\eta_{MN}$.

Let us compare the loop equation \eqref{0th_loopeq} with
the Hamiltonian constraint \eqref{HC}.
We find that, if we make the following identifications,
\begin{align}
\lambda &= \frac{R^4}{2 \alpha'^2},
\label{identify1}
\\
( 2 \pi^2 \kappa )^{-1/4}\epsilon &=  Y ,
\label{identify2}
\end{align}
the loop equation \eqref{0th_loopeq} can be regarded as the
Hamiltonian constraint \eqref{HC} acting on the Wilson loop up to
the $(\delta/\delta\eta)(\delta/\delta\ol{\zeta})$ term and
the omitted $\ldots$ terms in \eqref{0th_loopeq}.
The first identification \eqref{identify1} is standard in the
${\rm AdS}_5/{\rm CFT}_4$ correspondence \cite{Maldacena:1997re,
Gubser:1998bc,Witten:1998qj,Aharony:1999ti,D'Hoker:2002aw}.
On the other hand, the second identification \eqref{identify2} is
rather problematic and has no justification yet.
Roughly, it identifies the UV cutoff $\epsilon$ in SYM with the radial
coordinate $Y$ of ${\rm AdS}_5$.
This may look natural if we recall that $Y=0$ corresponds to the AdS
boundary \cite{Susskind:1998dq}, and might imply that we are forced to
consider only the strings on the AdS boundary.
Another possibility would be that we can treat finite $Y$ through the
relation \eqref{identify2} in the limit $\epsilon\to 0$ of removing
the UV cutoff by fine-tuning $\kappa$ in such a way that the LHS of
\eqref{identify2} is finite.
For this fine-tuning, the first term
$(3/8\pi)\epsilon\delta (s_1-s_1)/\sqrt{( \dot x (s_1))^2}$
in $\kappa$ \eqref{kappa} must be a finite quantity as we
mentioned below \eqref{bosondeltafree}, and it should be taken to
$1/\pi^2$.
This claims that the UV regularization of $\delta(s_1-s_1)$, namely,
the string worldsheet regularization, should be related with the
spacetime regularization specified by $\epsilon$.
In any case, justification of \eqref{identify2} is indispensable for
the identification of the Wilson loop with string field mentioned in
section 1.

\section{Loop equation in $\bm{\N = 4}$ SYM II:
linear and quadratic terms in fermionic coordinates}
\label{General}
In section \ref{Wilson44}, we have shown that our extended Wilson
loop operator \eqref{lowestW} satisfies the loop equation
\eqref{pre_loopeq2} with $\zeta=\eta=0$.
If we do not set $\zeta=\eta=0$ from outside, the $\delta$-term is no
longer proportional to the EOM.
For constructing the loop equation valid to higher powers in fermionic
coordinates, we must further modify the Wilson loop operator and/or
consider other types of functional derivatives.
In this section, we will take the former prescription
of modifying the Wilson loop and derive the loop equation valid to
terms quadratic in fermionic coordinates.
In appendix \ref{generaldiff}, we consider extending the
quadratic functional derivative $K_{\beta_1}$.
We find, however, that this does not change much the results of this
section.

Let us take following Wilson loop which is a generalization of
\eqref{lowestW}:
\begin{align}
 & W(C) \notag \\
 &=
 \Tr \Po\!
 \exp
 \biggl(
     i \int_0^l ds
     \left(
       \A_M (x(s),\zeta(s)) \dot X^M (s)
       +
       \ol{\Phi}(x(s),\zeta(s)) \dot \eta (s)
       +
       \dot {\ol{\zeta}} (s) \Omega(x(s),\zeta(s))
     \right)
   \biggr),
   \label{generalW}
\end{align}
with
\begin{align}
  \A_M
  &=
  A_M
  +
  a_1 \delta_\zeta A_M
  +
  a_2 \delta_\zeta^2 A_M
  +
  a_3 \delta_\zeta^3 A_M
  +
  \cdots, \label{generalA}\\
  \Phi_\alpha
  &=
  \Psi_\alpha
  +
  b_1 \delta_\zeta \Psi_\alpha
  +
  b_2 \delta_\zeta^2 \Psi_\alpha
  +
  b_3 \delta_\zeta^3 \Psi_\alpha
  +
  \cdots, \label{generalP}\\
  \Omega_\alpha
  &=
  i\bigl(
  c_3  \delta_\zeta A_M
  +
  \cdots
  \bigr)(\Gamma^M \zeta)_\alpha
  ,
  \label{O}
\end{align}
and consider $K_{\beta_1}W(C)$ with $K_{\beta_1}$ given by
\eqref{K_beta1} to quadratic order in $\zeta$ and $\eta$.
Operators \eqref{generalA}, \eqref{generalP} and \eqref{O}
are defined by using the SUSY transformation of SYM fields
and the parameters $a_n$, $b_n$ and $c_n$.%
\footnote{
The index $n$ denotes the power of $\zeta$ in the exponent of
\eqref{generalW}.
One might think it natural to introduce $A_M (\Gamma^M \zeta)_\alpha$
as the lowest order term in $\Omega_\alpha$. However, this operator
is excluded since it breaks the gauge invariance of the Wilson loop.
}
The coefficient $\beta_1$ in $K_{\beta_1}$ helps us to
distinguish contributions from
$(\delta/\delta X^M)^2$ and $\delta^2 /\delta\eta\delta\ol{\zeta}$.
Since the calculation of $K_{\beta_1}W(C)$ is lengthy and
complicated, we present it in appendix \ref{calc}.
The results are given by \eqref{finald^2W/dxdx} and
\eqref{finald^2W/dedz} using the notations \eqref{O_M}--\eqref{*Da},
and their explicit forms are found in appendices
\ref{partialdelta}--\ref{without}.
If we do not put $\zeta=\eta=0$ from outside, there appears the
$\dot\delta(s_1-s_2)$-term as well as the $\delta(s_1 - s_2)$-term in
$K_{\beta_1}W(C)$, and we have
\begin{align}
K_{\beta_1}\!W(C)
\!=\!\sum_{{\cal G},\widetilde{{\cal G}}}
\Tr\Po\bigl[{\cal G}_{s_1}\widetilde{{\cal G}}_{s_2}\W_0^l\bigr]
\!+\delta(s_1 - s_2)\Tr\bigl[\Op_{s_1} \W_{s_1}^{s_1+l}\bigr]
\!+ \dot\delta(s_1 - s_2)
 \Tr\bigl[{\cal Q}_{s_1}\W_{s_1}^{s_1+l}\bigr]
,  \label{del,partialdel}
\end{align}
where ${\cal G}$, $\widetilde{{\cal G}}$, $\Op$ and ${\cal Q}$ are
SYM operators.
We call the first, the second and the last term on the RHS of
\eqref{del,partialdel} \nd-term, $\delta$-term and $\dot\delta$-term,
respectively.

The parameters $a_n$, $b_n$ and $c_n$ should be determined from the
requirement that the operators $\Op$ and ${\cal Q}$ in the $\delta$-
and $\dot\delta$-terms in \eqref{del,partialdel} be proportional to
EOM.
In the following, we will summarize these operators and the
conditions for them to vanish modulo EOM
to terms quadratic in fermionic coordinates.
To this order of the fermionic coordinates, it is sufficient
to introduce $a_{1,2,3}$, $b_{1,2,3}$ and $c_3$.
Among these, we put $a_1=1$ by fixing the normalization of $\zeta$ as
we have already done in section \ref{Wilson44}.
Therefore we have to determine the six remaining parameters.

\subsection{Summary of operators and conditions}
\label{SummaryOpCond}

In this subsection, we summarize the operators $\cal O$ and ${\cal Q}$
appearing in \eqref{del,partialdel} and the corresponding conditions
to quadratic order in fermionic coordinates.
First, the operator ${\cal Q}$ in the $\dot\delta$-term is given simply
by (see \eqref{condPD})
\begin{equation}
{\cal Q}=9 \beta_1 b_2 \ol{\zeta} \DS \Psi .
\label{eq:calQ}
\end{equation}
This is already proportional to the EOM \eqref{fermioneq} and leads to
no conditions on the parameters.

Second, note that the operator $\Op$ in the $\delta$-term is given as
a sum of operators which are multiplied by one of
$(\dot X^M,\dot\zeta,\dot\eta)$.
We classify these operators by $(\dot X^M,\dot\zeta,\dot\eta)$ and the
power of $\zeta$ (the other fermionic coordinate $\eta$ appears in
$K_{\beta_1}W(C)$ only as $\dot\eta$). In the following we present each
operator and the corresponding condition.
Detailed calculations are given in appendix \ref{calc}, and we quote
only the results:
\begin{itemize}
\item  ${\dot  X}$-term \eqref{dotX}
\begin{align}
  \textrm{operator:} & \quad i
  \left(
    D^M F_{MN} - 2 \beta_1 \ol{\Psi} \Gamma_N \Psi
  \right) \dot X^N,  \\
  \textrm{condition:} &\quad
 \beta_1 = \frac{1}{2}. \label{beta1=1/2again}
\end{align}

\item $\dot\zeta$-term \eqref{dotz}
\begin{align}
  \textrm{operator:}& \quad - \dot {\ol{\zeta}} \DS \Psi,  \\
  \textrm{condition:}&\quad \textrm{none}.
\end{align}

\item $\dot \eta$-term \eqref{dote}
\begin{align}
  \textrm{operator:} & \quad
  i \dot {\ol{\eta}} \DS^2 \Psi
  -
  \frac{1 - \beta_1 b_1}{2}
  \left[
    F_{MN} , \ol{\Psi} \Gamma^{MN} \dot \eta
  \right],  \\
  \textrm{condition:} & \quad
  \beta_1 b_1 = 1.
\end{align}

\item $\zeta \dot X$-term \eqref{zdotX}
\begin{align}
  \textrm{operator:} & \quad
  \left(
    \ol{\zeta} \Gamma_M \DS^2 \Psi
    -
    \ol{\zeta} D_M \DS \Psi
  \right)\dot X^M  %\notag \\
   +
  \frac{i\beta_1 a_2}{2}
  \left[
     F_{NP}
     ,
    \ol{\Psi} \left\{ \Gamma_M , \Gamma^{NP} \right\} \zeta
  \right]  \dot X^M \notag \\
  & \qquad -
  \frac{i\left(1 - \beta_1 b_1\right)}{2}
  \left[
    F_{NP} , \ol{\zeta} \Gamma^{NP} \Gamma_M \Psi
  \right] \dot X^M, \\
  \textrm{conditions:} & \quad
  \beta_1 a_2 = 0, \quad
  \beta_1 b_1 = 1.
 \end{align}

\item $\zeta\dot\zeta$-term \eqref{zdotz}
\begin{align}
  \textrm{operator:} & \quad
    \left( 1 + 3 \beta_1 c_3 \right)
   \dot {\ol{\zeta}} \Gamma^M  \zeta \ol{\Psi} \Gamma_M \Psi,  \\
   \textrm{condition:} &  \quad
  1 + 3 \beta_1 c_3 = 0.
\end{align}

\item $\zeta \dot \eta$-term \eqref{zdotefinal}
\begin{align}
  \textrm{operator:} & \quad \notag \\
  &\hspace{-1cm} i
  \left[
    \ol{\zeta} \DS \Psi , \ol{\Psi} \dot \eta
  \right]
  - i \beta_1 b_2
  \Bigl(
  \dot {\ol{\eta}}\zeta\left\{\ol{\Psi},\DS\Psi\right\}
    +
  \left[
    \ol{\Psi} \Gamma_M \zeta
    ,
    \dot {\ol{\eta}} \Gamma^M \DS \Psi
  \right]
  -
  \left[
    \ol{\Psi} \dot \eta , \ol{\zeta} \DS \Psi
  \right]
  \Bigr) \notag \\
  &\hspace{-1cm} + b_1\left(
  1 - \beta_1 b_1
  \right)
  \left[F^M{}_N , F_{MP} \right] \ol{\zeta} \Gamma^{NP} \dot \eta
  -i
  D_N \left(
    b_1 D^M F_{MP}
    -
     2 \beta_1 b_2
    \ol{\Psi} \Gamma_P \Psi
  \right)
  \ol{\zeta} \Gamma^{NP} \dot \eta
  \notag \\
  &\hspace{-1cm}
  +2i
  \left(1 -  \beta_1 b_2 \right)
  \left[
    \ol{\zeta} \Gamma^M \Psi
    ,
    D_M \ol{\Psi} \dot \eta
  \right],  \\
  \textrm{conditions:} & \quad
   b_1\left(1 - \beta_1 b_1\right) = 0,\quad b_1 = 2  \beta_1 b_2,
  \quad \beta_1 b_2 = 1.
\end{align}

\item $\zeta \zeta \dot X$-term \eqref{condzetazetaX}
\begin{align}
  \textrm{operator:}\quad
  &a_2 D_Q
  \left(D^M F_{MP} - \ol{\Psi} \Gamma_P \Psi \right)
  \ol{\zeta} \Gamma_N \Gamma^{PQ} \zeta \dot X^N
  +
  \left[
    \ol{\zeta} \DS \Psi , \ol{\zeta} \Gamma_N \Psi
  \right]  \dot X^N
  \notag\\
  &+
  \left( a_2 - \beta_1 a_3 \right)
  \left(
     \left[
      \ol{\zeta} \Gamma_N \Psi , \ol{\zeta} \DS \Psi
    \right]
    +
    \left[
      \ol{\zeta} \Gamma_M \Psi ,
      \ol{\zeta} \Gamma_N \Gamma^M \DS \Psi
    \right]
  \right)\dot X^N \notag \\
  &
  + i a_2
  \left(
    1 - 2 \beta_1 b_1
  \right)
  \left[ F^M{}_P , F_{MQ}\right]
  \ol{\zeta} \Gamma_N \Gamma^{PQ} \zeta \dot X^N\notag \\
  &
  + i a_2
  \left(
    1  -\beta_1 b_1
  \right)
  \left[ F_{PQ} , F_{MN} \right]
  \ol{\zeta} \Gamma^M \Gamma^{PQ} \zeta \dot X^N\notag \\
  &
  +
  2
  \left( 1 - a_2 - \beta_1 b_2 \right)
  \left[
    \ol{\zeta} \Gamma^M \Psi , \ol{\zeta} \Gamma_N D_M  \Psi
  \right]\dot X^N \notag \\
  &
  -
  \left( 1 - \beta_1 \left( 2 a_3 + 2 b_2  + c_3  \right)\right)
  \left[
    \ol{\zeta} \Gamma^M \Psi , \ol{\zeta} \Gamma_M D_N \Psi
  \right] \dot X^N\notag \\
  &
  -
  \left( a_2  - \beta_1 \left( 3 a_3 + b_2 \right) \right)
  \left[
    \ol{\zeta} \Gamma_N \Gamma^{MP} \Psi , \ol{\zeta} \Gamma_M D_P \Psi
  \right]\dot X^N, \\
  \textrm{conditions:} \quad
  & a_2 ( 1 - 2 \beta_1 b_1)=0,  \quad
   a_2 ( 1 -  \beta_1 b_1)=0,  \quad
  1 - a_2 - \beta_1 b_2 =0,  \quad  \notag   \\
  &1 - \beta_1 ( 2 a_3 + 2 b_2 + c_3 )=0, \quad
  a_2 - \beta_1 ( 3 a_3 + b_2 ) =0.
\end{align}

\end{itemize}

\subsection{Solution to the conditions}

In the previous subsection, we obtained twelve conditions on six
parameters $a_{2,3}$, $b_{1,2,3}$ and $c_3$ (\eqref{beta1=1/2again}
is merely the repetition of the result \eqref{beta1=1/2} in section
\ref{Wilson44}).
This (apparently overdetermined) set of conditions can in fact be
consistently solved to give
\begin{align}
  &a_2 = 0 ,\quad a_3 = - \frac{2}{3}, \notag\\
  &b_1 = 2 ,\quad b_2 = 2,\quad b_3 = {\rm arbitrary},
 \label{valueofpara}\\
  &c_3 = -\frac{2}{3} . \notag
\end{align}
Note that $a_n$ and $b_n$ in \eqref{valueofpara} are different from
those for the finite SUSY transformation, $a_n=b_n=1/n!$.

Using above results, $K_{\beta_1=1/2}W(C)$ is now given by
\begin{align}
  &K_{\beta_1=\frac12} W(C)=\mbox{\nd-term} +
  9\,\dot\delta(s_1 - s_2)
  \Tr
  \bigl[
    (\ol{\zeta} \DS \Psi)_{s_1} \W_{s_1}^{s_1+l}
  \bigr] \notag \\
  &\quad +
  \delta(s_1 - s_2)
  \Tr
  \biggl[
  \biggl(
  i
  \left(
      D^M F_{MN} - \ol{\Psi} \Gamma_N \Psi
  \right)
  \dot X^N
  -\dot {\ol{\zeta}} \DS \Psi
  + i \dot {\ol{\eta}} \DS^2 \Psi
\notag\\
&\qquad  +
  \left(
    \ol{\zeta} \Gamma_M \DS^2 \Psi
    -
    \ol{\zeta} D_M \DS \Psi
  \right)
  \dot X^M
  - 2 i D_N
  \left(
    D^M F_{MP} - \ol{\Psi} \Gamma_P \Psi
  \right)
  \ol{\zeta} \Gamma^{NP} \dot \eta
  - i \dot {\ol{\eta}} \zeta
  \left\{ \ol{\Psi} , \DS \Psi \right\} \notag \\
  &\qquad - i
  \left[
    \ol{\Psi} \Gamma_M \zeta
    ,
    \dot {\ol{\eta}} \Gamma^M \DS \Psi
  \right]
  -
  \frac{1}{3}
  \left[
    \ol{\zeta}\Gamma_M \Psi
    ,
    \ol{\zeta}\Gamma^M \Gamma_N \DS \Psi
  \right] \dot X^N
  \biggr)_{s_1}
  \W_{s_1}^{s_1+l}
  \biggr] ,
  \label{delta_terms}
\end{align}
up to terms higher than quadratic in fermionic coordinates.
The expression of the \nd-term is found in appendix \ref{without}.
In appendix \ref{free}, we carry out the analysis of the most singular
and linear part of the RHS of \eqref{delta_terms}. This is an
extension of the analysis presented in section \ref{0thLeq} to the
quadratic order in the fermionic coordinates. The contribution from
the \nd-term is given by \eqref{nondeltafree}, and that from the
$\delta$- and $\dot\delta$-terms by \eqref{deltafree}.
We do not know whether whole of the most singular part,
including in particular its fermionic coordinate part, has an
interpretation as the Hamiltonian constraint of superstring on
${\rm AdS}_5\times{\rm S}^5$.
For this, we have to clarify the meaning of our fermionic coordinates
$\zeta$ and $\eta$ in the first-quantized superstring theory.

\section{Conclusion and discussions}
\label{Conclusion}

We have investigated the loop equation of the four-dimensional
$\N=4$ SYM.
We started with the Wilson loop operator introduced in
\cite{Rey:1998ik,Maldacena:1998im}, which contains six
scalar fields as well as the gauge field, and depends on extra six
bosonic coordinates $y^i(s)$ besides four-dimensional spacetime
coordinate $x^\mu(s)$.
We extended this Wilson loop to
include fermionic fields by introducing two fermionic coordinates
$\zeta(s)$ and $\eta(s)$: $\zeta(s)$ was introduced as the parameter
of the SUSY transformation, and $\eta(s)$ was needed by dimension
counting arguments of the quadratic functional derivative of the loop
equation.
In section \ref{Wilson44}, we derived the loop equation by putting
these fermionic coordinates equal to zero from outside.
We found that a rather simple functional derivative is sufficient to
derive the loop equation.
In section \ref{General} and appendix \ref{calc}, we extended our loop
equation to quadratic order in fermionic coordinates.
In deriving this loop equation, we introduced six free parameters
in the Wilson loop which should be fixed by requiring that the
$\delta$-terms be proportional to the EOM of $\N=4$ SYM.
This requirement leads to twelve conditions on the six parameters,
which is apparently overdetermined. However, we can consistently solve
this system of equations for the parameters.
We expect that there is some cleverer and concise derivation of the
loop equation valid to all powers of fermionic coordinates.
Understanding the meaning of the fermionic coordinate
$\eta$ would be important for this purpose.

We also extracted the most singular and linear part of our
loop equation to the lowest order in the gauge coupling constant.
There are two origins of such singular part, which we called \nd-term
and $\delta$-term in section \ref{0thLeq}.
It is interesting that the singular part from the \nd-term gets
ten-dimensionally covariant only after the contributions
from both the fermionic and the bosonic quadratic functional
derivatives are added.

Our original aim of studying the loop equation is to extract some
information about the string field theory on
${\rm AdS}_5 \times {\rm S}^5$ or pp-wave background.
Although we could not perform such an investigation yet in
this paper, we compared the most singular and linear part of our loop
equation with the Hamiltonian constraint of bosonic string on
${\rm AdS}_5\times {\rm S}^5$.
We found that these two equations take the same form if we identify
the UV cutoff $\epsilon$ of SYM and the radial coordinate $Y$
of ${\rm AdS}_5$.

However, there remain many problems to be clarified in the study of
the loop equation in $\N=4$ SYM: For example, establishing the loop
equation valid to higher orders in fermionic coordinates, and more
satisfactory analysis of the singular part of the loop equation.
The latter problem includes the treatment of the UV cutoff $\epsilon$
and $\delta(s_1-s_1)$, and the analysis beyond the expansion in SYM
coupling constant.
It is our future subject to resolve these problems to reach complete
understanding of the relation between the loop equation in $\N=4$ SYM
and the string field equation on ${\rm AdS}_5\times {\rm S}^5$.

\section*{Acknowledgments}
We would like to thank M.\ Asano, M.\ Fukuma, H.\ Itoyama, H.\ Kawai,
T.\ Kuroki, T.\ Morita, H.\ Shimada and A.\ Tsuchiya
for valuable discussions and comments.
The work of H.\ H.\ was supported in part by the Grant-in-Aid for
Scientific Research (C) No.\ 15540268 from Japan Society for the
Promotion of Science (JSPS).
The work of A.\ M.\ was supported in part by the Grant-in-Aid for the
21st Century COE "Center for Diversity and Universality in Physics"
from the Ministry of Education, Culture, Sports, Science and
Technology (MEXT) of Japan.

\section*{Appendix}

\appendix
\section{Notations and conventions} \label{notation}
In this appendix, we summarize our notations and conventions.
We use the following five sets of indices:
\begin{align}
  M,N  &= 0 ,\ldots , 9 ,
 \quad \mbox{(10-dim spacetime index)},\\
  \mu,\nu &= 0, \ldots ,3 ,
 \quad \mbox{(4-dim spacetime index)},\\
  i,j &= 1, \ldots , 6,
 \quad\mbox{(index for the scalars)},\\
  \alpha,\beta &= 1, \ldots ,32,
 \quad\mbox{(SO$(9,1)$ spinor index)},\\
  a,b &= 1,\ldots, N^2 - 1,
 \quad\mbox{(SU$(N)$ gauge index)}.
\end{align}
Our conventions for the flat metric tensors are
\begin{align}
\eta_{MN} = {\rm diag} (-1,1,\cdots,1),\qquad
\eta^{(4)}_{MN} = {\rm diag} (-1,1,1,1,0,\ldots,0).
\end{align}
The fields in four-dimensional $\N = 4$ SYM are given as follows:
\begin{align}
  A_\mu = A_\mu^a t^a &: \quad \textrm{gauge field}, \\
  A_{3+i} = A_{3+i}^a t^a &: \quad
  \textrm{six scalar fields}, \\
  \Psi_\alpha = \Psi_\alpha^a t^a &: \quad
  \textrm{16 component 10-dim Majorana-Weyl spinor},
\end{align}
where the generators $t^a$ of SU($N$) gauge group are hermitian matrices
normalized by
\begin{align}
  \Tr \left( t^a t^b \right) = \frac{1}{2} \delta^{ab}.
\end{align}
Note that we have gathered four four-dimensional Weyl spinors to
define one ten-dimensional Majorana-Weyl spinor $\Psi$.

The gamma matrices $\Gamma_M$ and
$\Gamma_{MN}=(1/2)\left[\Gamma_M,\Gamma_N\right]$
enjoy the following identities:
\begin{align}
  \Gamma_M\Gamma_N&=\Gamma_{MN}+\eta_{MN} ,\\
  \Gamma_M \Gamma^{PQ}
  &=
  \Gamma^{PQ} \Gamma_M
  +
  2 \delta_M{}^P \Gamma^Q
  -
  2 \delta_M{}^Q \Gamma^P,
  \label{G_MG^PQ} \\
  \Gamma_M \Gamma^{MN}
  &=
  - \Gamma^{MN} \Gamma_M
  =
  9 \Gamma^N,
  \label{G_MG^MN}\\
  \left[\Gamma_{MN} ,\Gamma_{PQ}\right]
  &=
  2\left( \eta_{MQ} \Gamma_{NP}
  -
   \eta_{MP} \Gamma_{NQ}
  -
   \eta_{NQ} \Gamma_{MP}
  +
  \eta_{NP} \Gamma_{MQ}\right).
  \label{[G,G]}
\end{align}
The Dirac conjugate of $\Psi$ is defined by
\begin{equation}
\ol{\Psi}_\alpha=\Psi_\beta C_{\beta\alpha} ,
\label{Diracconj}
\end{equation}
with the charge conjugation matrix $C$ satisfying
\begin{align}
  (C \Gamma^M C^{-1})_{\alpha \beta} = - (\Gamma^M)_{\beta \alpha}.
  \label{CC}
\end{align}
The following identity is often used in the calculations in appendix
\ref{calc} and also in showing the invariance of the action
\eqref{action} under the SUSY transformation \eqref{SUSY}:
\begin{align}
  \ol{\xi}_1 \Gamma^M \xi_2
  \,\ol{\xi}_3 \Gamma_M \xi_4
  +
  \ol{\xi}_1 \Gamma^M \xi_3
  \,\ol{\xi}_4 \Gamma_M \xi_2
  +
  \ol{\xi}_1 \Gamma^M \xi_4
  \,\ol{\xi}_2 \Gamma_M \xi_3
  =
  0,
  \label{powerful}
\end{align}
where $\xi_i$ ($i=1,\ldots,4$) are ten-dimensional Majorana-Weyl spinors
with a common chirality.

Finally, the anti-symmetrization $A_{[M}B_{N]}$ is defined by
\begin{equation}
A_{[M}B_{N]}=A_MB_N-A_NB_M .
\end{equation}

\section{Calculation of $\bm{K_{\beta_1}W(C)}$}
\label{calc}

In this appendix we present the explicit calculation of
$K_{\beta_1}W(C)$ for our Wilson loop operator \eqref{generalW}.
First, in \ref{functionaldiff} we give the expressions
\eqref{finald^2W/dxdx} and \eqref{finald^2W/dedz}  valid without
specifying the form of $\A_M$, $\Phi$ and $\Omega$.
Then in \ref{partialdelta}, \ref{withdelta} and \ref{without}, each
term of \eqref{finald^2W/dxdx} and \eqref{finald^2W/dedz}
is evaluated for $\A_M$, $\Phi$ and $\Omega$ given by \eqref{generalA},
\eqref{generalP} and \eqref{O}.

\subsection{Functional derivatives of Wilson loop}
\label{functionaldiff}

Let us consider the extended Wilson loop \eqref{generalW}.
The explicit expressions of $\A_M$ \eqref{generalA} , $\Phi$
\eqref{generalP} and $\Omega$ \eqref{O} are given by
\begin{align}
  \A_M &
  =
  A_M
  -
  i \ol{\zeta} \Gamma_M \Psi
  -
  a_2 \frac{i}{2} F_{NP} \ol{\zeta} \Gamma_M \Gamma^{NP} \zeta
  +
  a_3 \ol{\zeta} \Gamma_N D_P \Psi \ol{\zeta} \Gamma_M \Gamma^{NP}
  \zeta
  +
  \cdots,
  \label{exgenA}
  \\
  \Phi_\alpha &
=
  \Psi_\alpha
  +
  \left(
  b_1 \frac{1}{2} F_{MN}
  +
  b_2 i \ol{\zeta} \Gamma_M D_N \Psi
  +
  b_3
  \Upsilon_{MN} +\cdots
  \right)
  \left(\Gamma^{MN}\zeta \right)_\alpha ,
  \label{exgenP}
  \\
  \ol{\Phi}_\alpha
  &=
  \ol{\Psi}_\alpha
  -
  \left(
  b_1 \frac{1}{2} F_{MN}
  +
  b_2 i \ol{\zeta} \Gamma_M D_N \Psi
  +
  b_3
  \Upsilon_{MN} +\cdots
  \right)
  \left( \ol{\zeta} \Gamma^{MN} \right)_\alpha ,
  \label{exgenbP}
  \\
  \Omega_\alpha &
  =
  c_3 \ol{\zeta} \Gamma_M \Psi
  \left( \Gamma^M \zeta \right)_\alpha
  +
  \cdots,
  \label{exgenO}
\end{align}
where $\Upsilon_{MN}$ is defined by
\begin{equation}
  \Upsilon_{MN}
  =
  - i
  \left[
    \ol{\zeta} \Gamma_M \Psi
    ,
    \ol{\zeta} \Gamma_N \Psi
  \right]
  -
  \frac{i}{2}
  D_M F_{PQ} \ol{\zeta} \Gamma_N \Gamma^{PQ} \zeta,  \label{Upsilon}
\end{equation}
and dots denote terms with higher powers of fermionic coordinates.
We have put $a_1=1$ in \eqref{exgenA}.
Similarly to the derivation of \eqref{d/dx^muW}, we obtain
the following formulas for the functional derivatives of the Wilson
line $\W_{u_1}^{u_2}$:
\begin{align}
  \frac{\delta}{\delta X^M(s)} \W_{u_1}^{u_2} &=
  \W_{u_1}^s
  (\Op_M)_s
  \W_s^{u_2}
  + \W_{u_1}^{u_2} ( i \A_M)_{u_2} \delta(s - u_2)
  - ( i \A_M )_{u_1} \W_{u_1}^{u_2} \delta(s - u_1),
  \label{d/dX^MW}
   \\
  \frac{\delta}{\delta \eta_\alpha(s)} \W_{u_1}^{u_2} &=
  \W_{u_1}^s
  (\Op_{\eta_\alpha})_s
  \W_s^{u_2}
  - \W_{u_1}^{u_2} \bigl(i \ol{\Phi}_\alpha \bigr)_{u_2} \delta(s - u_2)
  + \bigl(i \ol{\Phi}_\alpha \bigr)_{u_1}
  \W_{u_1}^{u_2} \delta(s - u_1),
 \label{d/detaW}
 \\
  \frac{\delta}{\delta \ol{\zeta}_\alpha(s)} \W_{u_1}^{u_2} &=
  \W_{u_1}^s
  (\Op_{\bar \zeta_\alpha})_s
  \W_s^{u_2}
  + \W_{u_1}^{u_2} \bigl( i \Omega_\alpha \bigr)_{u_2} \delta(s - u_2)
  - \bigl( i \Omega_\alpha \bigr)_{u_1} \W_{u_1}^{u_2} \delta(s - u_1),
 \label{d/dolzetaW}
\end{align}
with ${\cal O}_M$, ${\cal O}_{\eta_\alpha}$ and
${\cal O}_{\bar\zeta_\alpha}$ defined by
\begin{align}
  {\cal O}_M
  &=
  i \F_{MN} \dot X^N
  +
  i \D_M \ol{\Phi} \dot \eta
  +
  i  \dot {\ol{\zeta}}_\alpha \F_{M \alpha}, \label{O_M}\\
  {\cal O}_{\eta_\alpha}
  &=
  i \D_N \ol{\Phi}_\alpha \dot X^N
  -
  \left\{
    i \ol{\Phi}_\alpha
    ,
    i \ol{\Phi}_\beta
  \right\}
  \dot \eta_\beta
  +
  i \dot {\ol{\zeta}}_\beta \D_\beta \ol{\Phi}_\alpha, \label{O_e}\\
  {\cal O}_{\bar \zeta_\alpha}
  & =
      i \F_{\alpha N} \dot X^N
    +
    i \D_\alpha \ol{\Phi} \dot \eta
    -
    i \dot {\ol{\zeta}}_\beta
    \F_{\alpha \beta}. \label{O_z}
\end{align}
Here we have introduced generalized field strengths $\F$ and covariant
derivatives $\D$ as follows:
\begin{align}
  \F_{MN}
  &=
  \partial_M \A_N
  -
  \partial_N \A_M
  +
  i \left[ \A_M , \A_N  \right], \label{*Fmn} \\
  \F_{M \alpha}
  &=
  -
  \F_{\alpha M}
  =
  \partial_M \Omega_\alpha
  -
  \partial_{\bar \zeta_\alpha} \A_M
  +
  i
  \left[
    \A_M
    ,
    \Omega_\alpha
  \right], \label{*Fma}\\
   \F_{\alpha \beta}
  &=
  \partial_{\bar \zeta_\alpha} \Omega_\beta
  +
  \partial_{\bar \zeta_\beta} \Omega_\alpha
  +
  i
  \left\{
    \Omega_\alpha
    ,
    \Omega_\beta
  \right\}, \label{*Fab}\\
  \D_M {\cal O}
  &=
  \partial_M {\cal O}
  +
  i
  \left[
    \A_M
    ,
    {\cal O}
  \right], \label{*Dm} \\
  \D_\alpha {\cal O}
  &=
  \partial_{\bar \zeta_\alpha} {\cal O}
  +
  \begin{cases}
  i \left[\Omega_\alpha,\cal O \right]&
  \Op
  \textrm{: bosonic, } \\
   i \left\{\Omega_\alpha,\cal O\right\}&
  \Op
  \textrm{: fermionic. } \\
  \end{cases}
  \label{*Da}
\end{align}
Using \eqref{d/dX^MW}, \eqref{d/detaW} and \eqref{d/dolzetaW} twice,
we get
\begin{align}
  &\frac{\delta}{\delta X^M(s_2)}\frac{\delta}{\delta X_M(s_1)}
  W(C)
 =
  \Tr\!
  \left[
    ( \Op_M )_{s_1}
    \W_{s_1}^{s_2}
    ( \Op^M )_{s_2}
    \W_{s_2}^{s_1+l}
  \right]
  +
   \delta(s_1-s_2)
  \Tr\!
  \left[
    (\D_M \Op^M)_{s_1}
     \W_{s_1}^{s_1+l}
  \right],
  \label{finald^2W/dxdx}
  \\
  &
  \frac{\delta}{\delta \eta_\alpha (s_2)}
  \frac{\delta}{\delta \ol{\zeta}_\alpha (s_1)}
  W(C)
  =
 -
  \Tr
  \left[
    ( \Op_{\bar \zeta_\alpha} )_{s_1}
    \W_{s_1}^{s_2}
    ( \Op_{\eta_\alpha} )_{s_2}
    \W_{s_2}^{s_1+l}
  \right]
  \notag\\
&\hspace{3cm}
  -
  \delta(s_1 - s_2)
  \Tr
  \left[
    \left\{
      i \ol{\Phi}_\alpha
      ,
      \Op_{\bar \zeta_\alpha}
    \right\}_{s_1}
    \W_{s_1}^{s_1+l}
  \right]
  +
  \dot\delta(s_1 - s_2)
  \Tr
  \left[
    \left(
      i \D_\alpha \ol{\Phi}_\alpha
    \right)_{s_1}
    \W_{s_1}^{s_1+l}
  \right] .
  \label{finald^2W/dedz}
\end{align}
In the rest of this appendix, we will evaluate each term in
\eqref{finald^2W/dxdx} and \eqref{finald^2W/dedz} using the concrete
expressions of $\A_M$, $\Phi$ and $\Omega$ given by
 \eqref{exgenA}--\eqref{exgenO}.

\subsection{$\dot \delta$-term}\label{partialdelta}

The $\dot\delta$-term appears only in \eqref{finald^2W/dedz} and is
given by
\begin{equation}
  \dot\delta(s_1 - s_2)
  \Tr
  \left[
    \left(
      i \D_\alpha \ol{\Phi}_\alpha
    \right)_{s_1}
    \W_{s_1}^{s_1+l}
  \right] .
\end{equation}
{}From \eqref{exgenbP} and \eqref{*Da}, we have
\begin{align}
  \D_\alpha \ol{\Phi}_\alpha &=
  \partial_{\bar \zeta_\alpha} \ol{\Phi}_\alpha
  +
  i \left\{ \Omega_\alpha , \ol{\Phi}_\alpha\right\} \notag \\
  &=
  -
  b_2 i
  \left( \Gamma_M D_N \Psi \right)_\alpha
  \left( \ol{\zeta} \Gamma^{MN} \right)_\alpha
  +
  i
  b_3
  \Bigl(
    \left[
      \left(\Gamma_M \Psi \right)_\alpha
      ,
      \ol{\zeta} \Gamma_N \Psi
    \right]
    -
    \left( M \leftrightarrow N \right)
    \Bigr)
    \left( \ol{\zeta} \Gamma^{MN} \right)_\alpha
    \notag \\
    &\qquad
    + b_3 \frac{i}{2}
    D_M F_{PQ}
    \left(
      \left\{
        \Gamma_N , \Gamma^{PQ}
      \right\}
      \zeta
    \right)_\alpha
    \left( \ol{\zeta} \Gamma^{MN} \right)_\alpha
  + i c_3
  \left\{
    \ol{\zeta} \Gamma_M \Psi \left( \Gamma^M \zeta \right)_\alpha
    ,
    \ol{\Psi}_\alpha
  \right\} + O(\zeta^3)\notag \\
  &= - 9 i b_2 \ol{\zeta} \DS \Psi + O(\zeta^3),
  \label{condPD}
\end{align}
where we have used \eqref{G_MG^PQ}, \eqref{G_MG^MN} and
\begin{equation}
  D_M F_{PQ} \ol{\zeta} \Gamma^M \Gamma^P \Gamma^Q \zeta = 0.
\label{cyclic}
\end{equation}
The last equation follows from the fact that
$\ol{\zeta} \Gamma^M \Gamma^N \Gamma^P \zeta$ is totally
anti-symmetric and hence is cyclically symmetric with respect to
$(M,N,P)$ and that  $F_{PQ}$ satisfies the Bianchi identity.

\subsection{$\delta$-term} \label{withdelta}

Let us calculate the $\delta$-terms in \eqref{finald^2W/dxdx} and
\eqref{finald^2W/dedz}:
\begin{align}
  &\frac{\delta}{\delta X^M(s_2)}\frac{\delta}{\delta X_M(s_1)}
  W(C)
  \ni
  \delta(s_1-s_2)
  \Tr
  \left[
    \D_M
    \bigl(
      i \F^M{}_N \dot X^N
      +
      i \D^M \ol{\Phi} \dot \eta
      +
      i \dot {\ol{\zeta}}_\alpha \F^M{}_\alpha
    \bigr)_{s_1}
    \W_{s_1}^{s_1+l}
  \right],
  \label{delta_d^2W/dxdx}
  \\
  &
  \frac{\delta}{\delta \eta_\alpha (s_2)}
  \frac{\delta}{\delta {\ol{\zeta}}_\alpha (s_1)}
  W(C)
  \ni
  -
  \delta(s_1 - s_2)
  \Tr
  \left[
    \bigl\{
      i \ol{\Phi}_\alpha
      ,
      i \F_{\alpha N} \dot X^N
    +
    i \D_\alpha \ol{\Phi} \dot \eta
    -
    i \dot {\ol{\zeta}}_\beta
    \F_{\alpha \beta}
    \big\}_{s_1}
    \W_{s_1}^{s_1+l}
  \right].
  \label{delta_d^2W/dedz}
\end{align}
Here we have used the symbol $\ni$ to indicate that the LHS contains
the RHS.
The field strengths and the covariant derivatives are defined by
\eqref{*Fmn}--\eqref{*Da} and their explicit expressions are as
follows:
\begin{align}
  \F_{MN}
  &=
  F_{MN}
  -
  i \ol{\zeta} \Gamma_{[N} D_{M]} \Psi
  -
  \frac{i a_2}{2}\ol{\zeta}\Gamma_{[N}\Gamma^{PQ}\zeta D_{M]}
  F_{PQ}
  -
  i
  \left[
    \ol{\zeta} \Gamma_{M} \Psi
    ,
    \ol{\zeta} \Gamma_{N} \Psi
  \right]
  + O(\zeta^3),
  \label{Fmn}
  \\
  \D_M \ol{\Phi}_\alpha
  &=
  D_M
  \left(
    \ol{\Psi}
    -
    \frac{b_1}{2} F_{NP} \ol{\zeta} \Gamma^{NP}
  \right)_\alpha
  +
  i
  \left[
    - i \ol{\zeta} \Gamma_M \Psi
    ,
    \ol{\Psi}_\alpha
  \right]
  +
  O(\zeta^2),
  \label{DmPa}
  \\
  \F_{\alpha M} &= - \F_{M \alpha} =
  - i \left( \Gamma_M \Psi \right)_\alpha
  - \frac{i a_2}{2} F_{NP}
  \left(
    \left\{ \Gamma_M , \Gamma^{NP} \right\} \zeta
  \right)_\alpha
  +
  a_3 \left( \Gamma_N D_P \Psi\right)_\alpha
  \ol{\zeta} \Gamma_M \Gamma^{NP} \zeta
  \notag \\
  &
  \qquad \qquad \qquad+
  a_3 \ol{\zeta} \Gamma_N D_P \Psi
  \left(
    \left\{
      \Gamma_M , \Gamma^{NP}
    \right\} \zeta
  \right)_\alpha
  -
    c_3
    \ol{\zeta} \Gamma_N D_M \Psi
    \left(\Gamma^N \zeta \right)_\alpha
    +
  O(\zeta^3),
  \label{Fan}
  \\
  \D_\alpha \ol{\Phi}_\beta
  &=
  -b_2 i ( \Gamma_M D_N \Psi )_\alpha
  ( \bar \zeta \Gamma^{MN} )_\beta
  -
  \left(
    \frac{b_1}{2} F_{MN}
    +
    b_2 i \bar \zeta \Gamma_M D_N \Psi
  \right)
  ( \Gamma^{MN} )_{\alpha\beta}
  +
  O (\zeta^2),
  \label{DaPb}
  \\
  \F_{\alpha \beta}
  &=
  c_3
  \left( \Gamma_M \Psi \right)_\alpha
  \left(\Gamma^M \zeta\right)_\beta
  +
  c_3
  \ol{\zeta} \Gamma_M \Psi
  \left(
    - \Gamma^M C^{-1}
  \right)_{\alpha \beta}
  +
  \left( \alpha \leftrightarrow \beta \right)
  +
  O (\zeta^2).
  \label{Fab}
\end{align}
Using \eqref{Fmn}--\eqref{Fab} we have the following expressions for
the ingredients of \eqref{delta_d^2W/dxdx} and \eqref{delta_d^2W/dedz}:
\begin{align}
  &i \D^M \F_{MN}  \dot X^N=
  i
  \biggl(
  D^M
   \Bigl(
   F_{MN}
   -
   i \ol{\zeta}  \Gamma_{[N} D_{M]} \Psi
      -
   \frac{ia_2}{2}
   \ol{\zeta} \Gamma_{[N} \Gamma^{PQ} \zeta
   D_{M]} F_{PQ}
   -
   i
   \left[
     \ol{\zeta} \Gamma_M \Psi
     ,
     \ol{\zeta} \Gamma_N \Psi
   \right]
   \Bigr)
  \notag \\
  & \qquad\qquad\qquad\quad
  +
  \left[
    \ol{\zeta} \Gamma^M \Psi
    ,
    F_{MN}
    -
    i \ol{\zeta} \Gamma_{[N} D_{M]} \Psi
  \right]
  +
  \frac{a_2}{2}\left[
     F_{PQ} \ol{\zeta} \Gamma^M \Gamma^{PQ} \zeta
    ,
    F_{MN}
  \right]
  \biggr) \dot X^N,
  \label{DmFmn}\\
  &i \D^2 \ol{\Phi} \dot \eta=
  i  D^2
  \left(
    \ol{\Psi} \dot \eta
    -
    \frac{b_1}{2} F_{MN} \ol{\zeta} \Gamma^{MN} \dot \eta
  \right)
  +
  iD^M
  \left[
    \ol{\zeta} \Gamma_M \Psi
    ,
    \ol{\Psi} \dot \eta
  \right]
  +
  i \left[
    \ol{\zeta} \Gamma^M \Psi
    ,
    D_M \ol{\Psi} \dot \eta
  \right],
  \label{D^2Pa}\\
  &i \dot {\ol{\zeta}}_\alpha \D^M \F_{M \alpha}=
  - D^M
  \left(
  \dot {\ol{\zeta}}\Gamma_M \Psi
  +
  \frac{a_2}{2} F_{NP}
  \dot {\ol{\zeta}}
  \left\{
      \Gamma_M , \Gamma^{NP}
    \right\} \zeta
  \right)
  -
  \left[
    \ol{\zeta}\Gamma^M \Psi
    ,
    \dot{\ol{\zeta}}\Gamma_M \Psi
  \right],
  \label{DmFma}\\
  &\bigl\{
    \ol{\Phi}_\alpha
    ,
    \F_{\alpha M} \dot X^M
  \bigr\}
  =
  \biggl\{
    \ol{\Psi}_\alpha
    ,
      - i \left( \Gamma_M \Psi \right)_\alpha
  - \frac{ia_2}{2} F_{NP}
  \left(
    \left\{ \Gamma_M , \Gamma^{NP} \right\} \zeta
  \right)_\alpha
  +
  a_3 \left( \Gamma_N D_P \Psi\right)_\alpha
  \ol{\zeta} \Gamma_M \Gamma^{NP} \zeta
  \notag \\
  &
  \hspace{4.5cm}+
  a_3 \ol{\zeta} \Gamma_N D_P \Psi
  \left(
    \left\{
      \Gamma_M , \Gamma^{NP}
    \right\} \zeta
  \right)_\alpha
  -
    c_3
   \ol{\zeta} \Gamma_N D_M\Psi
    \left(\Gamma^N \zeta \right)_\alpha
  \biggr\} \dot X^M
  \notag \\
  &\hspace{3.5cm}-
  \left\{
    \frac{b_1}{2} F_{QR} \left( \ol{\zeta} \Gamma^{QR}
    \right)_\alpha
    ,
    - i \left( \Gamma_M \Psi \right)_\alpha
  - \frac{i a_2}{2} F_{NP}
  \left(
    \left\{ \Gamma_M , \Gamma^{NP} \right\} \zeta
  \right)_\alpha
  \right\} \dot X^M\notag \\
  &\hspace{3.5cm}-
  \left\{
    b_2 i \ol{\zeta} \Gamma_N D_P \Psi \left(
      \ol{\zeta} \Gamma^{NP}
    \right)_\alpha
    ,
      - i \left( \Gamma_M \Psi \right)_\alpha
  \right\}  \dot X^M,
  \label{[PaFan]}\\
  &\left\{
    \ol{\Phi}_\alpha
    ,
    \D_\alpha \ol{\Phi} \dot \eta
  \right\}
  =
  \biggl\{
    \ol{\Psi}_\alpha
    ,
      -b_2 i ( \Gamma_M D_N \Psi )_\alpha
  \bar \zeta \Gamma^{MN} \dot \eta
  -
  \left(
    \frac{b_1}{2} F_{MN}
    +
    b_2 i \bar \zeta \Gamma_M D_N \Psi
  \right)
  ( \Gamma^{MN} \dot \eta )_{\alpha}
  \biggr\}
  \notag \\
  & \hspace{3.5cm} +
  \left\{
    - \frac{b_1}{2} F_{PQ} \left( \ol{\zeta} \Gamma^{PQ}
    \right)_\alpha
    ,
    -
  \frac{b_1}{2} F_{MN}
  \left(
    \Gamma^{MN} \dot \eta
  \right)_\alpha
  \right\},
  \label{[PaDaPe]}\\
  & -\bigl\{
    \ol{\Phi}_\alpha
    ,
    \dot {\ol{\zeta}}_\beta \F_{\alpha \beta}
  \bigr\}
  =
  -
  c_3
  \Bigl\{
    \ol{\Psi}_\alpha
    ,
    -
    \left( \Gamma_M \Psi \right)_\alpha
  \dot{\ol{\zeta}} \Gamma^M \zeta
  +
  2
  \ol{\zeta} \Gamma_M \Psi
  \bigl(
    \Gamma^M \dot \zeta
  \bigr)_\alpha
  +
  \dot {\ol{\zeta}} \Gamma_M \Psi
  \left(\Gamma^M \zeta\right)_\alpha
  \Bigr\}.
  \label{[PazFb]}
\end{align}
Here we have kept only terms at most quadratic in fermionic
coordinates.

Let us present the explicit form of the operator $\Op$ in the
$\delta$-term of $K_{\beta_1}W(C)$ (see \eqref{del,partialdel})
As we explained in section \ref{SummaryOpCond},
it is given as a sum of terms which are classified by
$(\dot X^M,\dot\zeta,\dot\eta)$ and the power of $\zeta$ multiplying
them.
In the following, terms without $\beta_1$ come from \eqref{delta_d^2W/dxdx}
(and hence from \eqref{DmFmn}, \eqref{D^2Pa} and \eqref{DmFma}),
while those multiplied by $\beta_1$ from \eqref{delta_d^2W/dedz}
(and hence from \eqref{[PaFan]}, \eqref{[PaDaPe]} and
\eqref{[PazFb]}):

\begin{itemize}
\item ${\dot  X}$-term\\
This term comes from \eqref{DmFmn} and \eqref{[PaFan]}:
\begin{align}
  i D^M F_{MN} \dot X^N
  +
  \beta_1
  \left\{ \ol{\Psi}
      ,
      - i \Gamma_M \Psi
    \right\}
    \dot X^M
  =
  i
  \left(
    D^M F_{MN} - 2 \beta_1 \ol \Psi \Gamma_N \Psi
  \right) \dot X^N. \label{dotX}
\end{align}

\item $\dot \zeta$-term\\
This term appears only in \eqref{DmFma}:
\begin{align}
  - \dot {\ol{\zeta}} \DS \Psi. \label{dotz}
\end{align}

\item  $\dot \eta$-term\\
This term comes from \eqref{D^2Pa} and \eqref{[PaDaPe]}:
\begin{align}
  &i D^2 \ol{\Psi} \dot \eta
  -
   \frac{\beta_1 b_1}{2}
  \left[
    \ol{\Psi}
    \Gamma^{MN} \dot \eta
    ,
    F_{MN}
  \right]
  =
  i \dot {\ol{\eta}} \DS^2 \Psi
  -
  \frac{1 - \beta_1 b_1}{2}
  \left[
    F_{MN} , \ol{\Psi} \Gamma^{MN} \dot \eta
  \right],  \label{dote}
\end{align}
where we have used
\begin{align}
  D^2 \Psi = \DS^2 \Psi - \frac{1}{2} \Gamma^{MN} D_{[M} D_{N]} \Psi,
  \label{D^2toDS^2}
\end{align}
and
\begin{align}
  D_{[M}D_{N]} \Op = i \left[ F_{MN} , \Op \right], \label{DDtoF}
\end{align}
for any $\Op$.

\item $\zeta \dot X$-term

The $O(\zeta^1)$ terms in \eqref{DmFmn} and \eqref{[PaFan]}
contribute to this term:
\begin{align}
  &i
  \left(
    - i \ol{\zeta} D^M
    \Gamma_{[N} D_{M]} \Psi
    +
    \left[ \ol{\zeta} \Gamma^M \Psi , F_{MN} \right]
  \right) \dot X^N \notag \\
  &\quad+
  \frac{\beta_1}{2}
  \Bigl(
    \left[
      \ol{\Psi}
        \left\{
          \Gamma_M , \Gamma^{NP}
        \right\}
        \zeta
      ,
      - ia_2
      F_{NP}
    \right]
  +
  i b_1
  \left[
    F_{NP}
    ,
      \ol{\zeta} \Gamma^{NP}
      \Gamma_M \Psi
  \right]
  \Bigr)\dot X^M
  \notag \\
  &=
  \left(
    \ol{\zeta} \Gamma_M \DS^2 \Psi
    -
    \ol{\zeta} D_M \DS \Psi
  \right)\dot X^M \notag \\
  &\quad +
  \frac{i\beta_1 a_2}{2}
  \left[
     F_{NP}
     ,
    \ol{\Psi} \left\{ \Gamma_M , \Gamma^{NP} \right\} \zeta
  \right]  \dot X^M
  -
  \frac{i\left( 1 - \beta_1 b_1 \right)}{2}
  \left[
    F_{NP} , \ol{\zeta} \Gamma^{NP} \Gamma_M \Psi
  \right] \dot X^M , \label{zdotX}
\end{align}
where we have used \eqref{D^2toDS^2}, \eqref{DDtoF} and
\eqref{G_MG^PQ}.

\item $\zeta \dot \zeta$-term

This term comes from \eqref{DmFma} and \eqref{[PazFb]}:
\begin{align}
  &
  -
  \frac{a_2}{2} D_M F_{NP}
  \dot {\ol{\zeta}} \left\{ \Gamma^M , \Gamma^{NP} \right\} \zeta
  -
  \left[
    \ol{\zeta} \Gamma^M \Psi ,
    \dot {\ol{\zeta}} \Gamma_M \Psi
  \right]
  \notag \\
  &-
  \beta_1 c_3
  \left(
    -
    \left\{
      \ol{\Psi}
      ,
      \Gamma_M \Psi
    \right\}
    \dot {\ol{\zeta}} \Gamma^M \zeta
    +
    2
    \bigl[
      \ol{\Psi} \Gamma^M \dot \zeta
      ,
      \ol{\zeta} \Gamma_M \Psi
    \bigr]
    +
    \bigl[
      \ol{\Psi} \Gamma^M \zeta
      ,
      \dot {\ol{\zeta}} \Gamma_M \Psi
    \bigr]
  \right)
  \notag \\
  &=
  \left( 1 + 3 \beta_1 c_3 \right)
  \dot {\ol{\zeta}} \Gamma^M  \zeta \ol{\Psi} \Gamma_M \Psi.
  \label{zdotz}
\end{align}
Here we have used
\begin{align}
  D_M F_{NP}
  \dot {\ol{\zeta}}
  \left\{ \Gamma^M , \Gamma^{NP} \right\}
  \zeta
  =
  D_M F_{NP}\,\frac{d}{ds}\!
  \left( \ol{\zeta} \Gamma^M \Gamma^{NP} \zeta
  \right)
  =0,
\end{align}
where the last equality is due to the same argument as for \eqref{cyclic}.
We have also used \eqref{powerful} to rewrite all other terms
into the form of the RHS.

\item $\zeta \dot \eta$-term

This term has contributions from \eqref{D^2Pa} and \eqref{[PaDaPe]}:
\begin{align}
  &    - \frac{i b_1}{2} D^2 F_{MN}
  \ol{\zeta} \Gamma^{MN}\dot \eta
  +
  i
  D^M
  \left[ \ol{\zeta} \Gamma_M \Psi, \ol{\Psi} \dot \eta \right]
  +
  i
  \left[ \ol{\zeta} \Gamma^M \Psi, D_M \ol{\Psi} \dot \eta \right]
  \notag \\
  &
  +
  \beta_1
  \biggl(
  - i b_2
  \Bigl(
    \left\{
      \ol{\Psi}
      ,
      \Gamma_M D_N \Psi
    \right\}
    \ol{\zeta} \Gamma^{MN} \dot \eta
    +
    \left[
      \ol \Psi \Gamma^{MN} \dot \eta
      ,
      \ol{\zeta} \Gamma_M D_N \Psi
    \right]
  \Bigr)
  \notag\\
  &\hspace{6cm}
  +
  \frac{b_1^2}{8}
  \left[ F_{PQ} , F_{MN}\right]
  \ol{\zeta}
  \left[ \Gamma^{PQ} , \Gamma^{MN}\right] \dot \eta
  \biggr). \label{zdote}
\end{align}
Let us make the following rewritings of the terms in \eqref{zdote}.
The first term is rewritten using
\begin{equation}
D^2F_{MN}=2i\left[F^P{}_M,F_{PN}\right]-D_{[M}D^PF_{N]P} ,
 \label{D^2F}
\end{equation}
which is obtained by covariant differentiating the Bianchi identity
and using \eqref{DDtoF}.
For the last term of \eqref{zdote} we use \eqref{[G,G]}.
The terms multiplied by $b_2$ are rewritten as follows:
\begin{align}
  &\left\{
    \ol{\Psi}
    ,
      \Gamma_M D_N \Psi
  \right\}
    \ol{\zeta} \Gamma^{MN} \dot \eta
  =
  D_N \left( \ol{\Psi} \Gamma_M \Psi \right) \ol{\zeta} \Gamma^{MN}
  \dot \eta ,\\
  &\left[
    \ol{\Psi}
      \Gamma^{MN} \dot \eta
    ,
    \ol{\zeta} \Gamma_M D_N \Psi
  \right]
  =
  \ol{\zeta}\Gamma^{MN} \dot \eta
  \zeta D_N \left( \ol{\Psi} \Gamma_M \Psi \right)
  +
  2
  \left[
    \ol{\zeta} \Gamma^M \Psi , D_M \ol{\Psi} \dot \eta
  \right] \notag \\
  &\hspace{3cm}+
  \dot{\ol{\eta}}\zeta\left\{\ol{\Psi},\DS\Psi\right\}
  +
  \left[
    \ol{\Psi} \Gamma_M \zeta
    ,
    \dot {\ol{\eta}} \Gamma^M \DS \Psi
  \right]
  -
  \left[
    \ol{\Psi} \dot \eta , \ol{\zeta} \DS \Psi
  \right] ,
\end{align}
where we have used \eqref{powerful} for the latter.
Gathering all the terms, we find that \eqref{zdote} is rewritten as
\begin{align}
   &
   i
  \left[
    \ol{\zeta} \DS \Psi , \ol{\Psi} \dot \eta
  \right]
  - i \beta_1 b_2
  \Bigl(
  \dot {\ol{\eta}}\zeta\left\{\ol{\Psi},\DS\Psi\right\}
  +
  \left[
    \ol{\Psi} \Gamma_M \zeta
    ,
    \dot {\ol{\eta}} \Gamma^M \DS \Psi
  \right]
  -
  \left[
    \ol{\Psi} \dot \eta , \ol{\zeta} \DS \Psi
  \right]
  \Bigr) \notag \\
  &+ \left(
  b_1 - \beta_1 b_1^2
  \right)
  \left[F^M\,_N , F_{MP} \right] \ol{\zeta} \Gamma^{NP} \dot \eta
  -i
  D_N \left(
    b_1 D^M F_{MP}
    -
     2 \beta_1 b_2
    \ol{\Psi} \Gamma_P \Psi
  \right)
  \ol{\zeta} \Gamma^{NP} \dot \eta
  \notag \\
  &
  +2i
  \left(1 -  \beta_1 b_2 \right)
  \left[
    \ol{\zeta} \Gamma^M \Psi
    ,
    D_M \ol \Psi \dot \eta
  \right] .
\label{zdotefinal}
\end{align}

\item $\zeta \zeta \dot X$-term

This term comes from \eqref{DmFmn} and \eqref{[PaFan]}:
\begin{align}
  &i
  \biggl(
  -
  \frac{i a_2}{2}
   \ol{\zeta} \Gamma_{[N} \Gamma^{PQ} \zeta
  D^M
  D_{M]}F_{PQ}
  +
  \frac{a_2}{2}
  \left[
    F_{PQ} , F_{MN}
  \right]
  \ol{\zeta} \Gamma^M \Gamma^{PQ} \zeta
  \notag \\
  &\qquad
  - i D^M
  \left[
    \ol{\zeta} \Gamma_M \Psi
    ,
    \ol{\zeta} \Gamma_N \Psi
  \right]
  - i
  \left[
    \ol{\zeta} \Gamma^M \Psi
    ,
    \ol{\zeta} \Gamma_{[N} D_{M]} \Psi
  \right]
  \biggr)\dot X^N
  \notag \\
  &+
  \beta_1
  \biggl(
  a_3
  \left\{
    \ol{\Psi}
    ,
    \Gamma_N D_P \Psi
    \right\}
    \ol{\zeta} \Gamma_M \Gamma^{NP} \zeta
  +
  a_3
  \left[
    \ol{\Psi}
    \left\{
        \Gamma_M , \Gamma^{NP}
      \right\} \zeta
    ,
    \ol{\zeta} \Gamma_N D_P \Psi
  \right]\notag \\
  &\qquad
  -
  c_3
  \left[
    \ol{\Psi}\Gamma^N \zeta
    ,
    \ol{\zeta} \Gamma_N D_M \Psi
  \right]
  +
  \frac{i a_2 b_1}{4}
  \left[
    F_{QR} , F_{NP}
  \right]
  \ol{\zeta} \Gamma^{QR}
  \left\{ \Gamma_M , \Gamma^{NP} \right\} \zeta \notag \\
  &\qquad
  - b_2
  \left[
    \ol{\zeta} \Gamma_N D_P \Psi
    ,
    \ol{\zeta} \Gamma^{NP} \Gamma_M \Psi
  \right]
  \biggr)\dot X^M. \label{zzX}
\end{align}
First we consider the three terms multiplied by $a_2$.
The sum of the first two $a_2$-terms is rewritten as follows:
\begin{align}
  &D^M
  \left(
    D_M F_{PQ} \ol{\zeta} \Gamma_N \Gamma^{PQ} \zeta
    -
    D_N F_{PQ} \ol{\zeta} \Gamma_M \Gamma^{PQ} \zeta
  \right) \dot X^N
  +i
  \left[
    F_{PQ} , F_{MN}
  \right]
  \ol{\zeta} \Gamma^M \Gamma^{PQ} \zeta \dot X^N
  \notag \\
  &=
  2 D_Q
  \left(
    D^M F_{MP}
    -
    \ol{\Psi} \Gamma_P \Psi
  \right)
  \ol{\zeta} \Gamma_N \Gamma^{QP} \zeta \dot X^N
  +
  2 \left\{
    \ol{\Psi}, \Gamma_P D_Q \Psi
  \right\}
  \ol{\zeta} \Gamma_N \Gamma^{QP} \zeta
  \dot X^N
  \notag\\
  &\qquad+
  2 i
  \left[
    F^M{}_P , F_{MQ}
  \right]
  \ol{\zeta} \Gamma_N \Gamma^{PQ} \zeta \dot X^N
  +
  2
  i
  \left[
    F_{PQ} , F_{MN}
  \right]
  \ol{\zeta} \Gamma^M \Gamma^{PQ} \zeta\dot X^N ,\label{boson1}
\end{align}
where we have used \eqref{DDtoF}, \eqref{D^2F}
and \eqref{cyclic}. Here we have added and subtracted the current
term:
\begin{equation}
  D_Q \left( \ol{\Psi} \Gamma_P \Psi \right)
   \ol{\zeta} \Gamma_N \Gamma^{QP} \zeta \dot X^N
  =
  \left\{
    \ol{\Psi}, \Gamma_P D_Q \Psi
  \right\}
  \ol{\zeta} \Gamma_N \Gamma^{QP} \zeta
  \dot X^N . \label{current}
\end{equation}
The remaining $a_2$-term which is multiplied by $\beta_1$ can be
rewritten as follows:
\begin{align}
  &
  \left[
    F_{QR} , F_{NP}
  \right]
  \ol{\zeta} \Gamma^{QR}
  \left\{ \Gamma_M , \Gamma^{NP} \right\} \zeta \dot X^M \notag \\
  &=
  -
    8
    \left[
    F_{QR} , F^Q\,_N
  \right]
  \ol{\zeta}
  \Gamma_M
  \Gamma^{RN}
  \zeta \dot X^M
  -
  4
  \left[
    F_{QR} , F_{PM}
  \right]
  \ol{\zeta} \Gamma^P
  \Gamma^{QR}
  \zeta \dot X^M
  ,
\end{align}
where we have used \eqref{G_MG^PQ} and \eqref{[G,G]}.

Next we proceed to the remaining terms with fermionic fields.
We adopt three operators of the following forms as the basis of
independent operators:
\begin{align}
  \left[
    \ol{\zeta} \Gamma^M \Psi
    ,
    \ol{\zeta} \Gamma_N D_M \Psi
  \right] \dot X^N
  ,\quad
  \left[
    \ol{\zeta} \Gamma^M \Psi
    ,
    \ol{\zeta} \Gamma_M D_N \Psi
  \right] \dot X^N
  ,\quad
  \left[
    \ol{\zeta} \Gamma_N \Gamma^{MP} \Psi
    ,
    \ol{\zeta}\Gamma_M D_P \Psi
  \right] \dot X^N,
\end{align}
and rewrite the terms in \eqref{zzX} in terms of this basis and terms
linear in EOM \eqref{fermioneq}.
Terms with $a_3$ are rewritten as follows:
\begin{align}
  &
  \left\{
    \ol{\Psi}
    ,
    \Gamma_N D_P \Psi
  \right\}
    \ol{\zeta} \Gamma_M \Gamma^{NP} \zeta
  \,\dot X^M
  \notag \\
  &=
  \left(
  \ol{\zeta} \Gamma_M \Gamma^P \Gamma^N \Psi^a
  D_P \ol{\Psi}^b \Gamma_N \zeta
  +
  \ol{\zeta} \Gamma_M \Gamma^P \Gamma^N D_P \Psi^b
  \ol{\zeta} \Gamma_N \Psi^a
  \right)\dot X^M
  [t^a,t^b]
  \notag \\
  &=
  \left[
    \ol{\zeta} \Gamma_M \Gamma^{NP} \Psi
    ,
    \ol{\zeta} \Gamma_N D_P \Psi
  \right] \dot X^M
  -
  \left[ \ol{\zeta} \Gamma_M \Psi , \ol{\zeta} \DS \Psi \right] \dot X^M
  \notag \\
  &\qquad-
  \left[
    \ol{\zeta} \Gamma_N  \Psi
    ,
    \ol{\zeta} \Gamma_M \Gamma^N \DS \Psi
  \right] \dot X^M
  + 2
  \left[
    \ol{\zeta} \Gamma^N \Psi , \ol{\zeta} \Gamma_M D_N \Psi
  \right]  \dot X^M, \label{a_3_1}
\end{align}
and
\begin{align}
  &\hspace{0cm}
  \left[
    \ol{\Psi}\left\{
        \Gamma_N , \Gamma^{MP}
      \right\} \zeta
    ,
    \ol{\zeta} \Gamma_M D_P \Psi
  \right]\dot X^N \notag \\
  &=
  2
  \Bigl(
   \left[
    \ol \zeta
      \Gamma_N \Gamma^{MP}
    \Psi
    ,
    \ol \zeta \Gamma_M D_P \Psi
  \right]
  -
   \left[
    \ol \zeta
    \Gamma^P
     \Psi
    ,
    \ol \zeta \Gamma_N D_P \Psi
  \right]
  +
   \left[
    \ol \zeta
    \Gamma^M
    \Psi
    ,
    \ol \zeta \Gamma_M D_N \Psi
  \right]
  \Bigr) \dot X^N , \label{a_3_2}
\end{align}
where we have used \eqref{powerful} for \eqref{a_3_1}, and
\eqref{G_MG^PQ} for \eqref{a_3_2}.
Note that \eqref{a_3_1} is just the current term \eqref{current}
(up to sign) and we also make the rewriting \eqref{a_3_1} for the
second term on the RHS of \eqref{boson1}.
For the $b_2$-term we have
\begin{align}
   & \left[
    \ol \zeta \Gamma_N D_P \Psi
    ,
    \ol \zeta \Gamma^{NP} \Gamma_M \Psi
  \right]  \dot X^M
  =
  \Bigl(
  -
  \left[
    \ol \zeta
    \Gamma_M \Gamma^{NP}
    \Psi
    ,
     \ol \zeta \Gamma_N D_P \Psi
  \right]
  \notag\\
  &\hspace{4cm}
  -2
  \left[
    \ol \zeta
    \Gamma^N
    \Psi
    ,
    \ol \zeta \Gamma_N D_M \Psi
  \right]
  +
  2
  \left[
    \ol \zeta
    \Gamma^P
    \Psi
    ,
     \ol \zeta \Gamma_M D_P \Psi
  \right]
  \Bigr)\dot X^M ,
\end{align}
where we have used \eqref{G_MG^PQ}.

Gathering all the terms we finally obtain the following expression of
\eqref{zzX}:
\begin{align}
  &
  a_2 D_Q
  \left(D^M F_{MP} - \ol \Psi \Gamma_P \Psi \right)
  \ol \zeta \Gamma_N \Gamma^{PQ} \zeta \dot X^N
  +
  \left[
    \ol \zeta \DS \Psi , \ol \zeta \Gamma_N \Psi
  \right]  \dot X^N
  \notag\\
  &+
  \left( a_2 - \beta_1 a_3 \right)
  \left(
     \left[
      \ol \zeta \Gamma_N \Psi , \ol \zeta \DS \Psi
    \right]
    +
    \left[
      \ol \zeta \Gamma_M \Psi ,
      \ol \zeta \Gamma_N \Gamma^M \DS \Psi
    \right]
  \right)\dot X^N \notag \\
  &
  +
  i a_2
  \left(
    1 - 2 \beta_1 b_1
  \right)
  \left[ F^M\,_P , F_{MQ}\right]
  \ol \zeta \Gamma_N \Gamma^{PQ} \zeta \dot X^N\notag \\
  &
  +
  i a_2
  \left(
    1  - \beta_1 b_1
  \right)
  \left[ F_{PQ} , F_{MN} \right]
  \ol \zeta \Gamma^M \Gamma^{PQ} \zeta \dot X^N\notag \\
  &
  +
  2
  \left( 1-  a_2 - \beta_1 b_2 \right)
  \left[
    \ol \zeta \Gamma^M \Psi , \ol \zeta \Gamma_N D_M  \Psi
  \right]\dot X^N \notag \\
  &
  -
  \left( 1 - \beta_1 \left( 2 a_3 + 2 b_2 + c_3 \right)\right)
  \left[
    \ol \zeta \Gamma^M \Psi , \ol \zeta \Gamma_M D_N \Psi
  \right] \dot X^N\notag \\
  &
  -
  \left( a_2 - \beta_1 \left( 3 a_3 + b_2 \right) \right)
  \left[
    \ol \zeta \Gamma_N \Gamma^{MP} \Psi , \ol \zeta \Gamma_M D_P \Psi
  \right]\dot X^N. \label{condzetazetaX}
\end{align}

\end{itemize}

\subsection{\nd-term} \label{without}

The \nd-terms without delta function are
\begin{align}
  &\frac{\delta}{\delta X_M(s_2)}\frac{\delta}{\delta X^M(s_1)}
  W(C) \notag \\
  &\ni
  \Tr
  \left[
    \left(
      i \F_{MN} \dot X^N
      +
      i D_M \ol \Phi \dot \eta
      +
      i \dot {\ol \zeta}_\alpha \F_{M \alpha}
    \right)_{s_1} \! \! \!
    \W_{s_1}^{s_2}
    \left(
      i \F^M\,_P \dot X^P
      +
      iD^M \ol \Phi \dot \eta
      +
      i \dot {\ol \zeta}_\beta \F^M\,_\beta
    \right)_{s_2} \! \! \!
    \W_{s_2}^{s_1+l}
  \right],
  \label{nodelta_d^2W/dxdx}\\
  &
  \frac{\delta}{\delta \eta_\alpha (s_2)}
  \frac{\delta}{\delta {\ol \zeta}_\alpha (s_1)}
  W(C) \ni
  -
  \Tr
  \biggl[
    \left(
      i \F_{\alpha N} \dot X^N
      +
      i \D_\alpha \ol \Phi \dot \eta
      -
      i \dot {\ol \zeta}_\beta
      \F_{\alpha \beta}
    \right)_{s_1} \!\!\!
    \W_{s_1}^{s_2} \notag \\
    &\hspace{5.2cm}\times
    \left(
      i \D_N \ol \Phi_\alpha \dot X^N
      -
      \left\{
        i \ol \Phi_\alpha
        ,
        i \ol \Phi_\beta
      \right\}
      \dot \eta_\beta
      +
      i \dot {\ol \zeta}_\beta \D_\beta \ol \Phi_\alpha
    \right)_{s_2} \!\!\!
    \W_{s_2}^{s_1+l}
  \biggr], \label{nodelta_d^2W/dedz}
\end{align}
with
\begin{align}
  i \F_{MN} \dot X^N
  &=
  i
  \left(
  F_{MN}
  -
  i \ol \zeta \Gamma_{[N} D_{M]} \Psi
  -
  \frac{i a_2}{2}
  \ol \zeta \Gamma_{[N} \Gamma^{PQ} \zeta
  D_{M]} F_{PQ}
\right) \dot X^N ,
  \\
  i\D_M \ol \Phi \dot \eta
  &=
  i D_M
  \left(
    \ol \Psi \dot \eta
    -
    \frac{b_1}{2} F_{NP} \ol \zeta \Gamma^{NP} \dot \eta
  \right) ,
  \\
  i \dot {\ol \zeta}_\alpha\F_{M \alpha} &=
  - \dot {\ol \zeta} \Gamma_M \Psi
  -
  \frac{a_2}{2} F_{NP}
    \dot {\ol \zeta}\left\{ \Gamma_M , \Gamma^{NP} \right\}
   \zeta ,
    \\
  i \F_{\alpha N} \dot X^N
  &=
  \biggl(
  \left( \Gamma_N \Psi \right)_\alpha
  +\frac{a_2}{2} F_{MP}
  \left(
    \left\{ \Gamma_N , \Gamma^{MP} \right\} \zeta
  \right)_\alpha
  +
  i a_3 \left( \Gamma_M D_P \Psi\right)_\alpha
  \ol \zeta \Gamma_N \Gamma^{MP} \zeta
  \notag \\
  &
  \qquad \qquad +
  i a_3 \ol \zeta \Gamma_M D_P \Psi
  \left(
    \left\{
      \Gamma_N , \Gamma^{MP}
    \right\} \zeta
  \right)_\alpha
  -
   i  c_3
    \ol \zeta \Gamma_M D_N \Psi
    \left(\Gamma^M \zeta \right)_\alpha
    \biggr) \dot X^N ,
  \\
  i \D_\alpha \ol \Phi \dot \eta
  &=
  -
  \frac{i b_1}{2} F_{MN}
  \left(
    \Gamma^{MN} \dot \eta
  \right)_\alpha
  +
  b_2
  \Bigl(
  \left(\Gamma_M D_N \Psi \right)_\alpha
  \ol \zeta \Gamma^{MN} \dot \eta
  +
  \ol \zeta \Gamma_M D_N \Psi
  \left(
    \Gamma^{MN} \dot \eta
  \right)_\alpha
  \Bigr) ,
  \\
  - i\dot {\ol \zeta}_\beta\F_{\alpha \beta}
  &=
  i
  c_3
  \left(\Gamma_M \Psi \right)_\alpha
  \dot {\ol \zeta} \Gamma^M \zeta
  -2 i
  c_3
  \ol \zeta \Gamma_M \Psi
  \left(
   \Gamma^M \dot \zeta
  \right)_\alpha
  -
  i c_3
  \dot {\ol \zeta} \Gamma_M \Psi
  \left(\Gamma^M \zeta \right)_\alpha ,
  \\
   i\D_N \ol \Phi_\alpha \dot X^N
  &=
  iD_N
  \left(
    \ol \Psi_\alpha
    -
    \left(
      \frac{b_1}{2} F_{MP}
      +
      b_2 i \ol \zeta \Gamma_M D_P \Psi
    \right)
    \left(
      \ol \zeta \Gamma^{MP}
    \right)_\alpha
  \right)\dot X^N , \\
  i \dot{\ol \zeta}_\beta \D_\beta \ol \Phi_\alpha
  &=
  b_2
  \dot {\ol \zeta} \Gamma_M D_N \Psi
  \left(\ol \zeta \Gamma^{MN} \right)_\alpha
  +
  \left(
  - i
  \frac{b_1}{2} F_{MN}
  +
  b_2
  \ol \zeta \Gamma_M D_N \Psi
  \right)
  \left(
    \dot{\ol \zeta} \Gamma^{MN}
  \right)_\alpha
  .
\end{align}
Here we have kept only terms which are at most quadratic in fermionic
coordinates and linear in SYM fields. The latter simplification is
because it is sufficient for our use in appendix \ref{free}.
We do not present the explicit form of $\{\ol \Phi, \ol \Phi\}$ in
\eqref{nodelta_d^2W/dedz} since it is already quadratic in fields.

\section{The most singular part of  the loop equation}
\label{free}

In this appendix we will extract the most singular and linear part of
the loop equation with non-zero fermionic coordinates.
We will consider lowest order terms in the coupling constant $g$.
This is an extension of the calculation in subsection \ref{LMS}.
First we consider the \nd-term in \ref{freewithoutpartdel},
and then in \ref{freedel} we consider the $\delta$- and the
$\dot\delta$-terms.

\subsection{\nd-term}
\label{freewithoutpartdel}

Let us consider the most singular part of \eqref{nodelta_d^2W/dxdx}
and \eqref{nodelta_d^2W/dedz} in the limit  $s_1 \to s_2$.
We use the formulas \eqref{DPP}, \eqref{FF} and
\begin{align}
  &\wick{1}
  {
    \partial_R \partial_M <1 A_N^a(x)
    \partial_P >1 A_Q^b
  }(\tilde x)\Bigr|_{x=\tilde x}
  =
  0 ,
  \label{aDDADA}
  \\
  &\wick{1}
  {
    <1 \Psi^a_\alpha(x)
    >1 {\overline \Psi}^b_\beta
  }(\tilde x)\Bigr|_{x=\tilde x}
   =
   \wick{1}
   {
     \partial_M
     <1 \Psi^a_\alpha(x)
     \partial_N
     >1 {\overline \Psi}^b_\beta
   }(\tilde x)\Bigr|_{x=\tilde x}
   =
   \wick{1}
   {
     \partial_N \partial_M
     <1 \Psi^a_\alpha(x)
     >1 {\overline \Psi}^b_\beta
   }(\tilde x)\Bigr|_{x=\tilde x}
   =
   0.\label{aDPDP}
\end{align}
The terms we consider are
\begin{equation}
K_{\beta_1}W(C) \ni
  \Tr \Po
  \Bigl[
    \Bigl(
 ({\cal O}_M)_{s_1}
 ({\cal O}^M)_{s_2}
 -
    \beta_1
    (\Op_{\bar \zeta_\alpha})_{s_1}
    (\Op_{\eta_\alpha})_{s_2}
  \Bigr)
    \W_0^l
  \Bigr],
\end{equation}
with
\begin{align}
  &
  (\Op_M)_{s_1}  (\Op^M)_{s_2}
  \notag \\
  &=
  \left(
    i F_{MN} \dot X^N
    -
    \frac{a_2}{2}
    F_{NP}
    \dot {\ol \zeta} \left\{ \Gamma_M , \Gamma^{NP} \right\}
    \zeta
    \right)_{s_1}
     \left(
    i F^M{}_Q \dot X^Q
    -
    \frac{a_2}{2}
    F_{QR}
    \dot {\ol \zeta} \left\{ \Gamma^M , \Gamma^{QR} \right\}
    \zeta
    \right)_{s_2}
    \notag \\
  &\qquad-
    \left(
      \ol \zeta \Gamma_{[N} D_{M]} \Psi \dot X^N
      +
      i D_M \ol \Psi \dot \eta
    \right)_{s_1}
    \bigl(
      \dot {\ol \zeta} \Gamma^M \Psi
    \bigr)_{s_2}
    -
  \bigl(
       \dot {\ol \zeta} \Gamma^M \Psi
    \bigr)_{s_1}
    \left(
      \ol \zeta \Gamma_{[N} D_{M]} \Psi \dot X^N
      +
      i D_M \ol \Psi \dot \eta
    \right)_{s_2} , \label{1}\\
  &
    (\Op_{\bar \zeta_\alpha})_{s_1}
    (\Op_{\eta_\alpha})_{s_2}
   \notag \\
  &=
    \left(
    \frac{a_2}{2}
    F_{MP}
    \left(
      \left\{ \Gamma_N , \Gamma^{MP} \right\}
      \zeta
    \right)_\alpha \dot X^N
    -
    \frac{i b_1}{2} F_{MN}
    \left( \Gamma^{MN} \dot \eta \right)_\alpha
    \right)_{s_1}
    \left(
      -\frac{ib_1}{2} F_{PQ}
      \bigl( \dot {\ol \zeta} \Gamma^{PQ} \bigr)_\alpha
    \right)_{s_2}  \notag \\
  &+
      \left(
      \left( \Gamma_N \Psi \right)_\alpha \dot X^N
    \right)_{s_1}
    \biggl(
      i D_N \ol \Psi_\alpha \dot X^N
      +
      b_2
      \Bigl(
      \dot {\ol \zeta} \Gamma_M D_N \Psi
      \left(\ol \zeta \Gamma^{MN} \right)_\alpha
      +
      \ol \zeta \Gamma_M D_N \Psi
      \bigl(
        \dot{\ol \zeta} \Gamma^{MN}
      \bigr)_\alpha
      \Bigr)
    \biggr)_{s_2}  \notag \\
  &+
    \left(
      i
      c_3
      \left(\Gamma_M \Psi \right)_\alpha
      \dot {\ol \zeta} \Gamma^M \zeta
      -2 i
      c_3
      \ol \zeta \Gamma_M \Psi
      \bigl(
        \Gamma^M \dot \zeta
      \bigr)_\alpha
      -
      i c_3
      \dot {\ol \zeta} \Gamma_M \Psi
      \left(\Gamma^M \zeta \right)_\alpha
    \right)_{s_1}
    \left(
      i D_N\ol \Psi_\alpha \dot X^N
    \right)_{s_2}.
   \label{2}
\end{align}
On the RHS's, we have kept only terms which are relevant to the present
analysis to quadratic order in fermionic coordinates.
We have also omitted terms like $(F^{MN})_{s_1} (D_P F_{QR})_{s_2}$,
$(\Psi)_{s_1}(\Psi)_{s_2}$,
$(D_M \Psi)_{s_1} (D_N \Psi)_{s_2}$ and
$(\Psi)_{s_1} (D_N D_P \Psi)_{s_2}$
 because these terms do not contribute to
the singular part we are considering
(see \eqref{aDDADA} and \eqref{aDPDP}).
The most singular part of each term on the RHS of \eqref{1} and
\eqref{2} is given as follows:
\begin{align}
  \textrm{The 1st term of \eqref{1}:} &\quad
  \frac{\lambda}{4 \pi^2}
  \frac{1}{\epsilon^4}
  \bigl(
    -4 \dot X^N \dot X_N - 8 \dot x^\mu \dot x_\mu
  \bigr)_{s_1},
  \\
  \textrm{The 2nd + 3rd terms of \eqref{1}:} &\quad
  \frac{\lambda}{4 \pi^2}
  \frac{1}{\epsilon^4}
  \bigl(
    -8 i \ol \zeta \Gamma_N \dot \zeta \dot X^N
    -
    16 i \ol \zeta \Gamma_\nu \dot \zeta \dot x^\nu
    +
    8 \dot {\ol \zeta} \dot \eta
  \bigr)_{s_1},
  \\
  \textrm{The 1st term of \eqref{2}:} &\quad
  \frac{\lambda}{4 \pi^2}
  \frac{1}{\epsilon^4}
  \Bigl(
    16
    i a_2 b_1
    \bigl(
      \dot {\ol \zeta} \Gamma_\mu \zeta \dot x^\mu
    -4 \dot {\ol \zeta} \Gamma_M \zeta \dot X^M
    \bigr)
      -
      36
      b_1^2
      \dot {\ol \zeta} \dot \eta
    \Bigr)_{s_1},
  \\
  \textrm{The 2nd term of \eqref{2}:} &\quad
  \frac{\lambda}{4 \pi^2}
  \frac{1}{\epsilon^4}
  \left(
    - 16  \dot x^\mu \dot x_\mu
  \right)_{s_1},
  \\
  \textrm{The 3rd term of \eqref{2}:} &\quad
  \frac{\lambda}{4 \pi^2}
  \frac{1}{\epsilon^4}
  \bigl(
  -i 24 c_3 \dot {\ol \zeta} \Gamma_\mu \zeta \dot x^\mu
  \bigr)_{s_1}.
\end{align}
Then summing all the contributions we get following expression for
the most singular part:
\begin{align}
  &
    ( \Op_M )_{s_1} ( \Op^M )_{s_2}
    -
  \beta_1
    (\Op_{\bar \zeta_\alpha})_{s_1}
    (\Op_{\eta_\alpha})_{s_2}
   \notag \\
  &\underset{s_1\to s_2}{\to}
  \frac{\lambda}{4\pi^2}
  \frac{1}{\epsilon^4}
  \Bigl(
  -4 \dot X^M \dot X_M
  -
  \left( 8 - 16 \beta_1 \right) \dot x^\mu \dot x_\mu
  +
  \left( 8 + 36 \beta_1 b_1^2 \right) \dot {\ol \zeta} \dot \eta
  \notag \\
  & \qquad
  +
  \left( 8 i +  64 i \beta_1 a_2 b_1  \right)
  \dot {\ol \zeta} \Gamma_N \zeta \dot X^N
  +
  \left(
    16 i
    - 16 i \beta_1 a_2 b_1
    + 24 \beta_1 i c_3
  \right)
  \dot {\ol \zeta} \Gamma_\mu \zeta \dot x^\mu
  \Bigr)_{s_1} .
\end{align}
Substituting the value of the parameters \eqref{valueofpara},
the most singular part in the \nd-term of $K_{\beta_1=1/2}W(C)$
is given by
\begin{align}
    \frac{\lambda}{4 \pi^2}
  \frac{1}{\epsilon^4}
  \left(
    - 4 \dot X^M \dot X_M
    +
    80 \dot {\ol \zeta} \dot \eta
    +
    8 i \dot {\ol \zeta} \Gamma_N \zeta \dot X^N
    +
    8 i \dot {\ol \zeta} \Gamma_\mu \zeta \dot x^\mu
  \right)_{s_1} W(C).
\label{nondeltafree}
\end{align}
Note that here we have considered only the lowest order terms in $g$.

\subsection{$\dot\delta$-term and $\delta$-term}
\label{freedel}

Next, let us consider the $\dot\delta$- and $\delta$-terms in
\eqref{delta_terms}, which consist of terms proportional to EOM.
For them we carry out the functional integration by parts with respect
to the SYM fields as we did in \eqref{intbyparts} to obtain
expressions quadratic in the Wilson loop (the zero-th order in the
fermionic coordinates is given by the last term of
\eqref{pre_loopeq2}).
Then we extract the most singular part of the $s$-integration which
comes from the region $s\sim s_1$ and is linear in Wilson loop.

First, let us consider the functional integration by parts.
Since we are interested in the terms which are at most quadratic in
fermionic coordinates in the final expression,
we can neglect a number of terms in \eqref{delta_terms}.
We obtain the following expression for the $\dot\delta$- and
$\delta$-terms in \eqref{delta_terms}:
\begin{align}
  9\,g^2
  \int_{s_1}^{s_1+l} ds&
  \,\ol \zeta_{s_1}
  (\Gamma_M \zeta \dot X^M - i \dot \eta  )_s
  \delta^{(4)} (x_s - x_{s_1})
  \Tr [t^a \W_{s_1}^s t^a \W_s^{s_1+l}]
   \dot\delta(s_1 - s_2)
  \notag \\
  + i g^2 \int_{s_1}^{s_1+l} ds
  &\biggl[
  \Bigl\{
    \dot {\ol \zeta}_{s_1}
  \bigl(
    i \Gamma_N \zeta\dot X^N
    +
    \dot \eta
  \bigr)_s
  -
    (\dot X^N)_{s_1}
    (\dot X_N)_s
  \Bigr\}
  \delta^{(4)} (x_s - x_{s_1})
    \Tr
    \left[
      t^a \W_{s_1}^s t^a \W_s^{s_1+l}
    \right]
    \notag \\
  & \qquad+ 2
  (\dot X^N)_{s_1}
  (\ol \zeta  \Gamma^M{}_N \dot \eta )_s
  \Tr
    \left[
      t^a \W_{s_1}^s D_M^s
      \left(
        \delta^{(4)} (x_s - x_{s_1}) t^a
      \right)
      \W_s^{s_1 + l}
    \right]
   \notag \\
  & \qquad+
  \Bigl\{
  2  ( \ol \zeta \Gamma^{MN} \dot \eta )_{s_1}
  ( \dot X_N )_s
  -
  \bigl(
  i \dot {\ol \eta} \Gamma^M
  + \dot X^P \ol \zeta \Gamma_P \Gamma^M
  -     \dot X^M \ol \zeta
  \bigr)_{s_1}
  (
    i \Gamma_N \zeta \dot X^N
    +
    \dot \eta
  )_s
 \Bigr\} \notag \\
  &
  \hspace{3cm}
  \times
    \Tr
    \left[
      D_M^{s_1}
      \left(
        \delta^{(4)} (x_s - x_{s_1}) t^a
      \right)
      \W_{s_1}^s t^a \W_s^{s_1+l}
    \right]
    \biggr] \delta(s_1 - s_2),
\label{afterFIBP}
\end{align}
where $D_M^s$, for example, denotes the covariant derivative with
respect to the coordinate $x(s)$.
Since we are interested in the lowest order terms in $g$,
we can replace covariant derivatives $D_M$ to partial derivatives
$\p_M$. After this simplification, \eqref{afterFIBP} is rewritten as
\begin{align}
  i g^2 \int_{s_1}^{s_1+l} ds
  \left(
    f_{(s_1,s_2)}(s) \delta^{(4)} (x(s) - x(s_1))
    +
    f_{(s_1,s_2)}^\mu (s) \partial_\mu^s \delta^{(4)} (x(s) - x(s_1))
  \right)
  \Tr
  \left[
    t^a \W_{s_1}^s t^a \W_s^{s_1+l}
    \right]
  , \label{deltaWW}
\end{align}
with
\begin{align}
  f_{(s_1,s_2)} (s)
  &=
  \Bigl\{
  - (\dot X^N)_{s_1}
  (\dot X_N )_s
  +
  \dot {\ol \zeta}_{s_1}
  \bigl(
    i \Gamma_N \zeta  \dot X^N
    +
    \dot \eta
  \bigr)_s
  \Bigr\}\delta(s_1 - s_2) \notag \\
  & \qquad \qquad
  - i\,9\, \ol \zeta_{s_1}
  \bigl(
  \Gamma_M \zeta \dot X^M - i \dot \eta
  \bigr)_s
  \dot\delta(s_1 - s_2), \\
  f_{(s_1,s_2)}^\mu (s)
  &=
  \Bigl\{
  2 (\dot X^N )_{s_1}
  ( \ol \zeta  \Gamma^\mu{}_N \dot \eta )_s
    -
    2 ( \ol \zeta \Gamma^{\mu N} \dot \eta )_{s_1}
    ( \dot X_N )_s
    \notag \\
    &\qquad \qquad
    +
    \bigl(
    i \dot {\ol \eta}
  \Gamma^\mu
    +
    \dot X^M  \ol \zeta \Gamma_M \Gamma^\mu
        -
    \dot X^\mu \ol \zeta
    \bigr)_{s_1}
  \bigl(
    i \Gamma_N \zeta  \dot X^N
    +
    \dot \eta
  \bigr)_s
  \Bigr\}
  \delta(s_1 - s_2).
\end{align}
Let us evaluate the most singular part of \eqref{deltaWW} which
arises from the region $s\sim s_1$ and is linear in Wilson loop.
Using the regularized delta function \eqref{regdelta}, we obtain
\begin{align}
  &\int ds f_{(s_1,s_2)}(s) \delta^{(4)} (x(s) - x(s_1))
  \Tr
  \left[
    t^a \W_{s_1}^s t^a \W_s^{s_1+l}
  \right]
  \notag \\
  & \qquad \quad\sim
  \frac{2i}{\pi^2}
  \int ds f_{(s_1,s_2)}(s_1)
  \frac{\epsilon^2}
  {\left((s-s_1)^2 (\dot x(s_1))^2 + \epsilon^2\right)^3}
  \Tr
  \left[
    t^a t^a \W_{s_1}^{s_1+l}
  \right]
  \notag \\
  & \qquad \quad =
  \frac{3i}{4\pi}
  \frac{f_{(s_1,s_2)}(s_1)}{\epsilon^4}
  \frac{\epsilon}{\sqrt{(\dot x(s_1))^2}}
  \frac{N}{2}
    W(C) ,
    \label{fdelta} \\
  &\int ds f_{(s_1,s_2)}^\mu(s) \partial_\mu^s \delta^{(4)} (x(s) - x(s_1))
  \Tr
  \left[
    t^a \W_{s_1}^s t^a \W_s^{s_1+l}
  \right] \notag \\
  &\qquad \quad \sim
  -
  \frac{12i}{\pi^2}\int ds
  \frac{\epsilon^2}
  {
    \left(
      (s-s_1)^2 ( \dot x(s_1) )^2
      +
      \epsilon^2
    \right)^4
  }
  \left(
    f_{(s_1,s_2)}^\mu(s_1) + (s-s_1) \partial_s f_{(s_1,s_2)}^\mu(s)
    \big|_{s=s_1}
  \right) \notag \\
  & \hspace{4cm}\times\left(
    (s-s_1) \dot x_\mu(s_1)
    +
    \frac{1}{2}
    (s-s_1)^2
    \ddot x_\mu (s_1)
  \right)
  \Tr
  \left[
    t^a t^a \W_{s_1}^{s_1+l}
  \right]
  \notag \\
  &\qquad \quad =
  - \frac{3i}{4 \pi}
  \frac{1}{\epsilon^4}
  \left\{
    \frac{1}{2}
    f_{(s_1,s_2)}^\mu (s_1) \ddot x_\mu (s_1)
    +
    \partial_s f_{(s_1,s_2)}^\mu(s) \big|_{s=s_1} \dot x_\mu (s_1)
  \right\}
  \frac{1}{( \dot x (s_1) )^2}
  \frac{\epsilon}{\sqrt{( \dot x(s_1))^2}}
  \frac{N}{2}
    W(C)
  . \label{dotfdelta}
\end{align}
In \eqref{dotfdelta}, we have neglected terms which arises from
the Taylor expansion of
$ \Tr  [t^a \W_{s_1}^s t^a \W_s^{s_1+l}] $.
This is because such terms, in general, have additional fields $A_M$
and $\Psi$ and do not contribute in the lowest order in $g$.

Finally, the most singular and linear part in the $\dot\delta$- and
$\delta$-terms of $K_{\beta_1=1/2}W(C)$ is given by
\begin{align}
   &\frac{3\lambda}{8\pi}
    \frac{1}{\epsilon^4}
   \frac{\epsilon}{\sqrt{( \dot x(s_1) )^2}}
   \biggl[ 9 (\ol \zeta \dot \eta )_{s_1}
   \dot\delta(s_1 - s_1)
   +
   \biggl\{
      \dot X^N \dot X_N
     -
     i \dot {\ol \zeta} \Gamma_N \zeta \dot X^N
     -
     \dot {\ol \zeta} \dot \eta
      + \frac{1}{(\dot x)^2}
    \biggl(
      2 \dot X^N \dot x^\mu \dot{\ol \zeta} \Gamma_{\mu N} \dot \eta
      \notag \\ & \qquad
      +
      2 \dot X^N \dot x^\mu \ol \zeta \Gamma_{\mu N} \ddot \eta
      -
      2 \ddot X^N \dot x^\mu \ol \zeta \Gamma_{\mu N} \dot \eta
      \notag
      -
      \dot X^N \dot x^\mu \dot {\ol \eta}
      \Gamma_\mu \Gamma_N
      \dot \zeta
      -
      \ddot X^N \dot x^\mu \dot{\ol \eta} \Gamma_\mu \Gamma_N \zeta
      \notag
      +
      i \dot x_\mu \dot {\ol \eta} \Gamma^\mu \ddot \eta
      \notag \\ & \qquad
      +
      i \dot X^M \dot x^\mu \dot X^N \ol \zeta
      \Gamma_M \Gamma_\mu \Gamma_N \dot \zeta
      + i \dot X^M \ddot X^N \dot x^\mu \ol \zeta
      \Gamma_M \Gamma_\mu \Gamma_N
      \zeta
      + \dot X^M \dot x^\mu \ol \zeta \Gamma_M \Gamma^\mu \ddot \eta
      -\frac{1}{2} \dot x^\mu \ddot x_\mu \ol \zeta \dot \eta
      \notag \\ & \hspace{6cm}
      - i \dot x^\mu \dot x_\mu \dot X^N \ol \zeta \Gamma_N \dot
      \zeta
      - \dot x^\mu \dot x_\mu \ol \zeta \ddot \eta
    \biggr)
    \biggr\}_{s_1} \delta(s_1 - s_1)
    \biggr]
      W(C)
     . \label{deltafree}
\end{align}
This reduces to our previous \eqref{bosondeltafree} if we put
$\zeta(s)=\eta(s)=0$.
The total of the most singular and linear part in
$K_{\beta_1=1/2}W(C)$ is the sum of \eqref{nondeltafree} and
\eqref{deltafree}.

\section{More general quadratic functional derivatives}
\label{generaldiff}

In sections \ref{Wilson44}--\ref{General}, we considered only the
simplest version of the quadratic functional derivative,
$K_{\beta_1}$ \eqref{K_beta1}.
In this section, we will consider more general quadratic functional
derivative and discuss their influence on the analysis given in
section \ref{General}.
The new functional derivatives we will consider are the following
five which are classified into type 1 and type 2:
\begin{align}
  &\textrm{type 1:}\quad
  \dot X^M
  \frac{\delta}{\delta \eta}
  \Gamma_M
  \frac{\delta}{\delta \dot {\ol \eta}} ,\quad
  \ol \zeta \Gamma^M \dot \zeta
  \frac{\delta}{\delta \eta}
  \Gamma_M
  \frac{\delta}{\delta \dot {\ol \eta}},\quad
  \frac{\delta}{\delta \eta} \Gamma_M \zeta
  \dot {\ol \zeta} \Gamma^M
  \frac{\delta}{\delta \dot {\ol \eta}}, \label{type1}\\
  &\textrm{type 2:}\quad
  \frac{\delta}{\delta \eta} \Gamma_M \zeta \frac{\delta}{\delta X^M},
  \quad
  \left(\frac{\delta}{\delta \eta}\Gamma_M \zeta \right)^2 .
  \label{type2}
\end{align}
All of these operators are of mass-dimension 2. Note that the type 1
operators contain $\delta/\delta\dot{\ol{\eta}}$.
As in $K_{\beta_1}$ \eqref{K_beta1}, one of the two functional
derivatives in each of the five operators in \eqref{type1} and
\eqref{type2} is at $s_1$ and the other at $s_2$, and we take the
limit $s_1\to s_2$ in the end.
As we will explain below, the type 2 operators \eqref{type2} have
ambiguity related to the choice of argument $s$ of $\zeta(s)$
multiplying them even after taking the limit $s_1 \to s_2$.
For this reason, we will not consider the type 2 operators in detail.
On the other hand, the type 1 operators \eqref{type1} are free from
such ambiguity.

\subsection{Type 1 operators}

Let us generalize $K_{\beta_1}$ to the following
$K_{\{\beta\}}$ obtained by
adding the type 1 operators:
\begin{align}
K_{\{\beta\}}&=\frac{\delta}{\delta X^M(s_2)}
  \frac{\delta}{\delta X_M(s_1)}
  +
  \beta_1
  \frac{\delta}{\delta \eta(s_2)}
  \frac{\delta}{\delta \ol \zeta(s_1)}
  +
  \beta_2
  \dot X^M (s_2)
  \frac{\delta}{\delta \eta (s_2)}
  \Gamma_M
  \frac{\delta}{\delta \dot {\ol \eta}(s_1)}
  \notag \\
  &\qquad+
  \beta_3
  \ol \zeta (s_2)
  \Gamma^M
  \dot \zeta (s_1)
  \frac{\delta}{\delta \eta(s_2)}
  \Gamma_M
  \frac{\delta}{\delta \dot {\ol \eta}(s_1)}
  +
  \beta_4
  \frac{\delta}{\delta \eta (s_2)}
  \Gamma^M
  \zeta(s_2)
  \dot {\ol \zeta}(s_1)
  \Gamma_M
  \frac{\delta}{\delta \dot {\ol \eta}(s_1)}
  .
\label{genaralK}
\end{align}
In the above type 1 operators, we have chosen deliberately the
argument $s$ of $\dot X^M(s)$ and $\zeta(s)$ multiplying them.
However, the choice of these arguments does not affect the following
discussion.
We will repeat the analysis of section \ref{General} by taking the
quadratic functional derivative $K_{\{\beta\}}$ \eqref{genaralK}
and the Wilson loop $W(C)$ \eqref{generalW}. Namely, we will determine
the coefficients $\beta_{1,2,3,4}$ in $K_{\{\beta\}}$ and $a_{2,3}$,
$b_{1,2,3}$ and $c_3$ in $W(C)$ from the requirement that the
$\delta$-term in $K_{\{\beta\}}W(C)$ be proportional to EOM.

The type 1 operators \eqref{type1} contain differentiation with
respect to $\dot {\ol \eta}$:
\begin{align}
  \frac{\delta}{\delta \dot {\ol \eta }(s)}
  W(C)
  &=
  \Tr
  \left[
    i \Phi_{s}
    \W_s^{s+l}
  \right].
\end{align}
Using this we have
\begin{align}
  &\dot X^M(s_2) \frac{\delta}{\delta \eta(s_2)}
  \Gamma_M
  \frac{\delta}{\delta \dot {\ol \eta}(s_1)}
  W(C)
  =
  - \Tr
  \left[
    (i ( \Gamma_M \Phi )_\alpha)_{s_1}
    \W_{s_1}^{s_2}
    (\Op_{\eta_\alpha})_{s_2}
    \W_{s_2}^{s_1+l}
  \right]
  \dot X^M(s_2)
  \notag \\
  &\hspace{6cm}
  -
  \delta (s_1 - s_2)\Tr
  \left[
    \left\{
      i \ol \Phi
      ,
      i \Gamma_M \Phi
    \right\}_{s_1} \!\!
    \W_{s_1}^{s_1+l}
  \right]
  \dot X^M(s_2),
  \label{D5}
\\
  &\ol \zeta (s_2) \Gamma^M \dot \zeta(s_1)
  \frac{\delta}{\delta \eta(s_2)}
  \Gamma_M
  \frac{\delta}{\delta \dot {\ol \eta}(s_1)}
  W(C)
  =
    - \Tr
  \left[
    (i ( \Gamma_M \Phi )_\alpha)_{s_1}
    \W_{s_1}^{s_2}
    (\Op_{\eta_\alpha})_{s_2}
    \W_{s_2}^{s_1+l}
  \right]
  \bar \zeta (s_2) \Gamma^M \dot \zeta (s_1)
  \notag \\
  & \hspace{5.1cm}
  -
  \delta (s_1 - s_2)
  \Tr
  \left[
    \left\{
      i \ol \Phi
      ,
      i \Gamma_M \Phi
    \right\}_{s_1} \!\!
    \W_{s_1}^{s_1+l}
  \right]
  \bar \zeta (s_2) \Gamma^M \dot \zeta (s_1)
  ,
  \label{D6}
\\
  &\frac{\delta}{\delta \eta (s_2)}
  \Gamma^M
  \zeta (s_2)
  \dot {\ol \zeta} (s_1)
  \Gamma_M
  \frac{\delta}{\delta \dot {\ol \eta}(s_1)}
  W(C)
  =
  \Tr
  \left[
    ( i \dot {\ol \zeta} \Gamma_M \Phi )_{s_1}
    \W_{s_1}^{s_2}
    ({\cal O}_{\eta} \Gamma^M \zeta)_{s_2}
    \W_{s_2}^{s_1+l}
  \right] \notag \\
  &\hspace{7cm}
  +
  \delta(s_1 - s_2)
  \Tr
  \left[
    \bigl[
      i \dot {\ol \zeta} \Gamma_M \Phi
      ,
      i \ol \Phi \Gamma^M \zeta
    \bigr]_{s_1} \!\!
    \W_{s_1}^{s_1+l}
  \right]
  ,
  \label{D7}
\end{align}
where $\Op_\eta \Gamma^M \zeta$  is the abbreviation of
$\Op_{\eta_\alpha} ( \Gamma^M \zeta )_\alpha $.
Following the same steps as in section \ref{General},
we obtain the following conditions on the parameters:
\begin{itemize}

\item {$ \dot X$}-term
\begin{equation}
  2 ( \beta_1 + i \beta_2 ) = 1,
\label{acondI}
\end{equation}

\item {$ \dot \zeta$}-term
\begin{equation}
  \textrm{none} ,
\end{equation}

\item $\dot\eta$-term
\begin{equation}
\beta_1 b_1 = 1,
\end{equation}

\item $\zeta\dot X$-term
\begin{equation}
  \beta_1 a_2 = 0 ,
  \quad
  1 - \beta_1 b_1 - 2 i \beta_2 b_1 = 0,
\end{equation}

\item $\zeta\dot\zeta$-term
\begin{equation}
  1 + 3 \beta_1 c_3 - 2\beta_3 + \beta_4 = 0,
\end{equation}

\item $\zeta\dot\eta$-term
\begin{equation}
  b_1\left(1 - \beta_1 b_1\right) = 0,
  \quad
  b_1 = 2 \beta_1 b_2,
  \quad
  \beta_1 b_2  = 1,
\end{equation}

\item $\zeta\zeta\dot X$-term
\begin{align}
  &i a_2 - 2 i \beta_1 a_2 b_1 + \beta_2 b_1^2  = 0
  , \quad
  i a_2 - i \beta_1 a_2 b_1+ \beta_2 b_1^2 = 0
  , \quad
  1 -  a_2 - \beta_1 b_2 - 2 i \beta_2 b_2  = 0
  , \notag \\
  & 1 - \beta_1 ( 2 a_3 + 2 b_2 + c_3  )
  - 4 i \beta_2 b_2 = 0
  ,\quad
  a_2 - \beta_1 ( 3 a_3 + b_2 ) - 2 i \beta_2 b_2  = 0.
\label{acondF}
\end{align}

\end{itemize}
The set of equations \eqref{acondI}--\eqref{acondF} can again be
consistently solved to give
\begin{equation}
\beta_1  = \frac{1}{2}, \quad
  \beta_2  = 0 , \quad
  \beta_3  = \frac{1}{2}\beta_4 ,
\end{equation}
and
\begin{align}
  &a_2  = 0 ,\quad
  a_3  = - \frac{2}{3}, \notag\\
  &b_1  = 2,\quad
  b_2  = 2,\quad
  b_3  = {\rm arbitrary}, \\
  &c_3  = -\frac{2}{3}. \notag
\end{align}
Namely, the values of the ``old parameters'' remain the same as
before: \eqref{beta1=1/2} and \eqref{valueofpara}.
Thus we have found that our new quadratic functional derivative
$K_{\{\beta\}}$ does not change essentially the results of section
\ref{General}.

\subsection{Type 2 operators}
Next, let us consider the type 2 operators \eqref{type2}.
Their action on the Wilson loop \eqref{generalW} is given by
\begin{align}
  &\frac{\delta}{\delta \eta(s_2)}
  \Gamma^M
  \zeta (s_2)
  \frac{\delta}{\delta X^M (s_1)}
  W(C)
  =
  \Tr
  \left[
    (\Op_M )_{s_1}
    \W_{s_1}^{s_2}
    (\Op_{\eta}  \Gamma^M \zeta)_{s_2}
    \W_{s_2}^{s_1+l}
  \right]
 \notag \\
  &\qquad
  +
  \delta (s_1 - s_2)
  \Tr
  \left[
    \left[
      \Op_M
      ,
      i \ol \Phi
      \Gamma^M
      \zeta
    \right]_{s_1}
    \W_{s_1}^{s_1 + l}
  \right]
  - i
  \dot\delta(s_1 - s_2)
  \Tr
  \left[
    \bigl(\D_M \ol \Phi \bigr)_{s_1} \Gamma^M \zeta_{s_2} \W_{s_1}^{s_1+l}
  \right]
  , \label{d^2W/dedx_z}
 \\
  &\frac{\delta}{\delta\eta(s_2)}
    \Gamma_M
    \zeta(s_2)
     \frac{\delta}{\delta\eta(s_1)}
    \Gamma^M
    \zeta(s_1)
   W(C)
  =
  \Tr
  \left[
    \bigl( \Op_\eta \Gamma^M \zeta \bigr)_{s_1}
    \W_{s_1}^{s_2}
    \bigl( \Op_\eta \Gamma_M \zeta \bigr)_{s_2}
    \W_{s_2}^{s_1+l}
  \right]
 \notag \\
  &
  +
  \delta(s_1-s_2)
  \Tr \!
  \left[
    \left[
      \Op_{\eta} \Gamma^M \zeta
      ,
      i \ol \Phi \Gamma_M \zeta
    \right]_{s_1} \!\!
    \W_{s_1}^{s_1+l}
  \right]
  +
  \dot\delta(s_1-s_2)
  \Tr \!
  \Bigl[
    \bigl[
      i ( \ol \Phi \Gamma^M \zeta )_{s_1}
      ,
      i \ol \Phi_{s_1} \Gamma_M \zeta_{s_2}
    \bigr]
    \W_{s_1}^{s_1+l}
  \Bigr]
  .
   \label{d^2W/de^2_z^2}
\end{align}
In the above type 2 operators, we have taken particular choices of the
arguments $s$ of $\zeta(s)$ multiplying them.
Let us consider the Taylor expansion of the $\dot\delta$-terms in
\eqref{d^2W/dedx_z} and \eqref{d^2W/de^2_z^2} with respect to $s_2$
around $s_1$ by using the formula:
\begin{align}
  f(s_1,s_2) \dot\delta( s_1 - s_2)
  &=
  \Bigl(
    f(s_1,s_1) +
    (s_2 - s_1) \partial_s f(s_1,s)\bigr|_{s=s_1} +
    \cdots
  \Bigr)
  \dot\delta(s_1 - s_2) \notag \\
  &=
  f(s_1,s_1) \dot\delta(s_1-s_2)
  +
  \partial_s f(s_1,s)\bigr|_{s=s_1} \delta(s_1 - s_2),
  \label{fdotdelta}
\end{align}
which follows from
\begin{align}
  s\,\dot\delta(s) = -\delta(s),\qquad
  s^n\dot\delta(s)=0\quad(n\ge 2).
\end{align}
We have
\begin{align}
  &\bigl(\D_M \ol \Phi\bigr)_{s_1}
  \Gamma^M \zeta_{s_2} \dot \delta(s_1 - s_2)
  =
  \left(D_M \ol \Psi  \Gamma^M \zeta\right)_{s_1}
  \dot\delta(s_1-s_2) \notag \\
  &\qquad\qquad\qquad +
  \Bigl(
  D_M \ol \Psi \Gamma^M \dot \zeta
  -
  \frac{b_1}{2}
  D_M F_{NP} \ol \zeta \Gamma^{NP} \Gamma^M \dot \zeta
  +
  \bigl[
    \ol \zeta \Gamma_M \Psi
    ,
    \ol \Psi \Gamma^M \dot \zeta
  \bigr]
  \Bigr)_{s_1}
  \delta(s_1-s_2) ,
\label{Taylor1}
\\
     &\bigl[
      i (\ol \Phi \Gamma^M \zeta )_{s_1}
      ,
      i\ol \Phi_{s_1} \Gamma_M \zeta_{s_2}
    \bigr] \dot\delta(s_1-s_2)
    =0\times\dot\delta(s_1-s_2)+
    \bigl[
      i \ol\Psi \Gamma^M \zeta
      ,
      i \ol \Psi \Gamma_M \dot \zeta
    \bigr]_{s_1}
    \delta(s_1 - s_2),
\label{Taylor2}
\end{align}
where we have kept only terms at most quadratic in fermionic
coordinates.
{}From \eqref{Taylor1} and \eqref{Taylor2}, we find the
followings. First, as in section \ref{General}, the $\dot\delta$-terms
on the RHS of \eqref{Taylor1} and \eqref{Taylor2} are already
proportional to the EOM (or equal to zero).
Second, the Taylor expansion of the $\dot\delta$-terms in
\eqref{d^2W/dedx_z} and \eqref{d^2W/de^2_z^2} give additional
contributions to the $\delta$-terms.
However, these new $\delta$-terms have ambiguities depending on the
choice of the arguments $s$ of $\zeta(s)$ multiplying the type 2
operators. For example, if we had adopted $s_1$ as the arguments of
all $\zeta(s)$ differently from \eqref{d^2W/dedx_z} and
\eqref{d^2W/de^2_z^2}, we would have obtained no additional
$\delta$-terms at all.
Therefore, the conditions determining the parameters depend on the
choice of the arguments of $\zeta(s)$.
For this reason, we do not consider the type 2 operators seriously
in this paper.
Note, however, that this kind of ambiguities do not appear in the case
of the type 1 operators since there are no $\dot\delta$-terms in
\eqref{D5}--\eqref{D7}.

\end{document}